\documentclass{aa}  

\usepackage{graphicx}
\usepackage{txfonts}
\usepackage{subcaption}         
\usepackage{lscape}             
\usepackage{placeins}  
                                
\usepackage{natbib,twoopt} 
\usepackage{float}
\usepackage[breaklinks=true]{hyperref} 
\bibpunct{(}{)}{;}{a}{}{,}             
\makeatletter
  \newcommandtwoopt{\citeads}[3][][]{\href{http://adsabs.harvard.edu/abs/#3}%
    {\def\hyper@linkstart##1##2{}%
     \let\hyper@linkend\@empty\citealp[#1][#2]{#3}}}
  \newcommandtwoopt{\citepads}[3][][]{\href{http://adsabs.harvard.edu/abs/#3}%
    {\def\hyper@linkstart##1##2{}%
     \let\hyper@linkend\@empty\citep[#1][#2]{#3}}}
  \newcommandtwoopt{\citetads}[3][][]{\href{http://adsabs.harvard.edu/abs/#3}%
    {\def\hyper@linkstart##1##2{}%
     \let\hyper@linkend\@empty\citet[#1][#2]{#3}}}
  \newcommandtwoopt{\citeyearads}[3][][]%
    {\href{http://adsabs.harvard.edu/abs/#3}
    {\def\hyper@linkstart##1##2{}%
     \let\hyper@linkend\@empty\citeyear[#1][#2]{#3}}}
\makeatother

\usepackage{epsfig}
\usepackage{txfonts}
\begin{document} 

\title{Jetted subgalactic-size radio sources in merging galaxies}

\subtitle{A jet redirection scenario}
   \author{C. Stanghellini \inst{1}
          \and M. Orienti \inst{1}
\and C. Spingola \inst{1}
\and A. Zanichelli \inst{1}
\and D. Dallacasa \inst{1,2}
\and P. Cassaro \inst{3}
\and C. P. O'Dea \inst{4}
\and S. A. Baum \inst {4}
\and  M. P\'erez-Torres \inst{5,6}
          }

\institute{INAF $-$ Istituto di Radioastronomia, via Gobetti 101, I-40129 Bologna, Italy  \\
              \email{carlo.stanghellini@inaf.it}
         \and
Dipartimento di Fisica e Astronomia, Universit\`a di Bologna, via Gobetti 93/2, I-40129 Bologna, Italy
\and
INAF $-$ Istituto di Radioastronomia,  Stazione di Noto, Contrada Renna
Bassa, I-96017 Noto, Italy
\and
Department of Physics \& Astronomy, University of Manitoba, 30A Sifton Rd., Winnipeg, MB R3T 2N2, Canada
\and
Consejo Superior de Investigaciones Científicas (CSIC) - Instituto de Astrofísica de Andalucía (IAA), Glorieta de la Astronomía s/n, 18008 Granada, Spain
\and 
School of Sciences, European University Cyprus, Diogenes street, Engomi, 1516 Nicosia, Cyprus
}
\titlerunning{JSS radio sources in merging galaxies - A jet redirection scenario}
\authorrunning{C. Stanghellini et al.}
   \date{\today}

  \abstract
     {The long-standing question concerning jetted subgalactic-size (JSS) radio sources is whether they will evolve into large radio galaxies,  die before escaping the host galaxy, or remain indefinitely confined to their compact size.  }
   {Our main goal is to propose a scenario that explains the relative number of JSS radio sources and their general properties.}
   {We studied the parsec-scale radio morphology of a complete sample of 21 objects using Very Long Baseline Array (VLBA) observations at various frequencies and analyzed the morphological characteristics of their optical hosts.}
   {Many of these radio sources exhibit radio morphologies consistent with transverse motions of their bright edges and are found in dynamically disturbed galaxies. VLBA images suggest the possible presence of large-angle, short-period precessing jets, and an orbital motion of the radio-loud active galactic nucleus (AGN) in a dual or binary system. The majority of JSS radio sources appear to be in systems in different stages of their merging evolution.}
   {We propose a scenario where rapid jet redirection, through precession or orbital motion, prevents the jet from penetrating the interstellar medium (ISM) sufficiently to escape the host galaxy. Most JSS radio sources remain compact due to their occurrence in merging galaxies.  }

   \keywords{Galaxies: active -- Galaxies: jets -- Galaxies: interactions -- Radio continuum: galaxies -- Techniques: high angular resolution -- Techniques: interferometric
               }

   \maketitle
%

\section{Introduction}
A significant fraction of extragalactic radio sources (10-30\%, depending on the flux density limit and observing frequency of the radio surveys) are intrinsically compact, with projected sizes smaller than those of their host galaxy. 
The simplest explanation for their compactness is that they represent the initial phase of radio activity in the active galactic nucleus (AGN) phenomenon and will eventually evolve into classical Fanaroff-Riley (FR) radio sources \citep{FR74}. However, if we assume that radio sources grow at a steady speed for most of their lifetime, their number is too large to support the ``youth scenario,'' unless we invoke a strong luminosity evolution \citep{1995A&A...302..317F, 1997AJ....113..148O}.

The alternative ``frustration'' scenario suggests that these radio sources are confined within the inner regions of the host galaxy by a dense interstellar medium (ISM) for an extended or even indefinite time \citep{1984AJ.....89....5V}. However, the existence of such a dense environment has remained elusive \citep{2000A&A...358..499F,2005ApJ...632..110S,2006MNRAS.367..928V}. Even when an interaction between jets and the ISM is observed, the amount of gas appears insufficient to confine the radio source indefinitely \citep{2004A&A...424..119M,2005A&A...436..493L,2006A&A...447..481L,2013Sci...341.1082M,2021A&A...647A..63S}.

The ``short-lived'' scenario is another alternative, proposing that most of these AGNs will cease radio activity before they can expand beyond the host galaxy \citep{1994cers.conf...17R,1997ApJ...487L.135R,2009ApJ...698..840C,2020MNRAS.499.1340O,2024ApJ...961..241K, 2024ApJ...961..240K}. In particular, it has been suggested that the majority of Compact Symmetric Objects (CSOs) are transient objects formed by a tidal disruption event of giant stars. These events involve an energy budget sufficient to turn on the radio source only for a very short time \citep{2024ApJ...961..240K,2024ApJ...961..241K,2024ApJ...961..242R,2024MNRAS.528.6302S}. 

Given the intrinsically small size of these radio sources, the emitting region undergoes synchrotron self-absorption. This generates a distinctive peak in the radio spectrum, which provides an effective tool for their selection \citep{1983A&A...123..107G,1998PASP..110..493O,2000MNRAS.319..445S}. Alternatively, the spectral peak may be caused by free-free absorption from ionized gas in front of the emitting region (e.g. \citealt{1997ApJ...485..112B,2015AJ....149...74T}).  A combination of both effects has been found in some objects (e.g. \citealt{2008A&A...479..409O}).

Historically, different names and acronyms have been used depending on where the radio peak occurred.
Compact Steep Spectrum (CSS) radio sources have been selected from low-frequency surveys as having a steep radio spectrum and a size smaller than 20 kpc \citep{1990A&A...231..333F,1995A&A...302..317F}.
GHz Peaked Spectrum (GPS) radio sources have been selected based on their convex radio spectrum peaking around 1 GHz, and generally have sizes not exceeding $\sim$1 kpc  \citep{1991ApJ...380...66O,1997AJ....113..148O,1998A&AS..131..303S}.
High Frequency Peakers (HFP) extend the GPS class to radio sources peaking around or above 5 GHz, with even smaller sizes \citep{2000A&A...363..887D,2009AN....330..223S}.
GPS radio sources and HFPs are included in the class of Peaked Sources (PS) in the extended review by \citet{2021A&ARv..29....3O}.
CSOs with a linear size (LS)<1 kpc, and middle-sized symmetric objects (MSOs) with 1 kpc < LS < 20 kpc, largely overlap with GPS and CSS radio sources, but are selected according to their radio structure rather than their spectral properties \citep{1994ApJ...432L..87W,1996ApJ...460..634R,1996ApJ...460..612R}.

Each of these selection criteria has its own limitations. Samples selected solely on the basis of spectral properties have been found to include a significant fraction of blazars, which contaminate the sample of intrinsically compact sources, exhibiting core-jet morphology and/or significant radio variability \citep{2003PASA...20..118S,2005A&A...435..839T,2007A&A...475..813O,2010MNRAS.408.1075O,2008A&A...479..409O}. Although it is true that some genuine compact sources may have the jet aligned with the line of sight, their spectra should generally be flat. The convex spectral shape seen in blazars is likely produced by a dominant component in a relativistic beamed jet. Although these blazars are interesting objects themselves, they constitute a distinct population from genuine compact radio galaxies.
Conversely, CSOs might exclude objects that lack clear two-sided structures for various reasons, but still belong to the same class of jetted radio sources embedded within their host galaxy.

To define a homogeneous class that includes all of the above categories based on the intrinsic property of interest, we introduce the class of jetted subgalactic-size (JSS) radio sources, meaning radio-loud AGNs with an intrinsic size of the radio emission that does not extend outside the host galaxy. Thus, the class of JSS radio sources includes all CSO/MSOs and CSS/PS sources, with the exclusion of contaminating blazars. 

JSS radio sources may have jets, lobes, hot spots, and cores, similar to large FR radio galaxies, but on a smaller scale. Not all of these components are necessarily visible, and central cores, in particular, appear rather elusive. 
The morphology is often irregular and the two sides of the radio source may differ significantly in size and brightness \citep{2003PASA...20...50S,2008A&A...487..865R,2013MNRAS.433..147D,2018A&ARv..26....4M}. These sources are weakly polarized, with fractional polarization increasing and Rotation Measure (RM) decreasing with size \citep{1998A&AS..131..303S,2001A&A...377..377S,2003PASA...20...12C,2004A&A...427..465F,2008A&A...487..865R}. The irregular morphology and depolarization suggest that the radio source is embedded in an ionized and inhomogeneous environment, which becomes clumpier toward the center of the galaxy. 
Faint and diffuse radio emission is occasionally detected on arcsecond to arcminute scales \citep{1990A&A...232...19B,2003PASA...20...16M,2005A&A...443..891S,2012A&A...544A..34P} and is often interpreted as relic emission from past AGN activity.

The optical hosts of JSS radio sources are generally galaxies, whereas the great majority of quasars found in samples of PS sources are contaminating objects that appear bright and compact due to beaming and projection effects \citep{2003PASA...20..118S}. 
JSS radio sources have been detected in X-rays only with the advent of sensitive X-ray telescopes. It is not clear whether the X-ray emission is of thermal origin (attributed to hot shocked gas or gas falling into the accretion disk) or of nonthermal origin (synchrotron or inverse Compton scattering) coming from the base of the jet \citep{2010ApJ...715.1071O,2017ApJ...851...87O,2019ApJ...884..166S}. Recent works, including $\gamma$-ray studies, suggest the presence of high-energy emission of nonthermal origin, possibly associated with the jet and/or lobes \citep{2021MNRAS.507.4564P,2023ApJ...948...81S,2024A&A...684A..65B}.

In this work, we present a radio morphological study at parsec scale for a complete sample of JSS radio sources, with additional considerations on the morphology of their optical hosts.  We also propose a novel scenario to explain the compactness and the excess number of JSS radio sources.

Throughout this paper, we assume the following cosmological parameters: H$_0$ =
70 km s$^{-1}$ Mpc$^{-1}$, $\Omega_M$ = 0.27, and $\Omega_\Lambda$ = 0.73, in a flat Universe.
Reported values from the literature have been adjusted, if necessary, to align with this cosmology.
The spectral index $\alpha$ is defined as S$(\nu) \propto \nu^{-\alpha}$. 

\section{The sample}
Many samples of JSS radio sources have been derived from various radio surveys, as listed in \cite{2021A&ARv..29....3O}.
In this paper, we focus on the 1 Jy complete sample of GPS radio sources selected by \cite{1998A&AS..131..303S}, to define a complete sample of JSS radio sources. In subsequent papers, we will expand the size range to include both larger and smaller scales, incorporating sources selected from current samples of CSS and HFP radio sources. 

As noted in the previous section, samples based on spectral properties may include contaminating blazars. Therefore, we refined the original 1 Jy sample, retaining only objects that exhibit emission on both sides of a putative center of activity, which may appear as a flat-spectrum compact component or be undetected due to sensitivity limitations.

The final complete sample of genuine JSS radio sources, listed in Table \ref{tablist}, consists of 21 radio sources with a flux density S>1 Jy at 5 GHz, declination $\delta>-25^\circ$, galactic latitude $\lvert b \rvert>10^\circ$, a radio spectrum peaking between 0.4 and 6 GHz, and a spectral index above the turnover frequency $\alpha > 0.5$. The selected radio sources are optically identified with galaxies, with a few exceptions.

\section{The data}

\subsection{Very Long Baseline Array (VLBA) observations and archival data}

The radio morphologies of the objects in our sample, with angular sizes ranging from a few to a few hundred mas (e.g. \citet{1997A&A...325..943S,1999A&AS..134..309S,2001A&A...377..377S}, are best studied in detail using very long baseline interferometry (VLBI) techniques.
We performed VLBA observations of 14 out of the 21 JSS radio sources in the sample (projects BS085 on 27 Mar, 30 Apr, 13 May, and 9 Jun 2001, and BO030 on 23 Mar and 27 Apr 2008). To complete the morphological study of the entire sample, we supplemented our observations with reprocessed archival data for the remaining sources.

Observations from projects BS085 and BO030 are highlighted in bold in Table \ref{tablist}. Although VLBI information is already available for many objects in the literature, we chose to present images for all objects in our complete sample, as additional important details are revealed in both proprietary and reprocessed archival data.  In a few cases, we concatenated data from different projects to improve the UV-coverage, as variability in flux density and structure is negligible. In Table \ref{tabproj} we list the archival VLBA data considered in this work.

Calibration of the VLBA data has been carried out using the Astronomical Image Processing System (AIPS; \citealt{1990apaa.conf..125G}) following standard procedures. The calibrated visibilities were then exported to the DIFMAP software package \citep{1997ASPC..125...77S} for self-calibration and imaging. The final images were imported back into AIPS for data analysis and to produce the figures shown in Appendix A (Figs. A.1 to A.21).

Table B.1 in Appendix B lists the flux densities calculated by using the IMSTAT task in AIPS, except for some compact components, where the flux densities were derived using the MODELFIT task in DIFMAP. This approach was used to better isolate their contribution from the surrounding diffuse emission.
Flux density errors for all components listed in Table \ref{comp} are dominated by calibration uncertainties and are conservatively estimated at 10\%, excluding the systematic contribution of missing extended flux density due to unsampled spatial frequencies.
The limitations in UV-coverage and image sensitivity prevented the creation of reliable spectral index maps. Consequently, the spectral index was calculated only for compact components or by integrating the flux density over isolated extended components. Even so, high-frequency images often fail to recover the full flux density of extended structures, leading to an artificially steepened spectrum.
Despite this, Table \ref{comp} includes the spectral index between the two highest available frequencies, which should be regarded as rough estimates and, for extended structures, upper limits. These values may still prove useful for identifying components with flatter spectral indices, potentially linked to a radio core. 

\subsection{Optical images}

We collected images of the optical hosts of our radio sources to review their optical morphology and intergalactic environment. We selected the best available images from various datasets, including the Sloan Digital Sky Survey (SDSS; \citealt{2022ApJS..259...35A}), the Pan-STARRS survey \citep{2016arXiv161205560C}, and reprocessed observations from the Nordic Optical Telescope (NOT) as reported by \cite{1993ApJS...88....1S}. When stacking images from different filters enhanced detail, we used the resulting composite images. The optical images, along with their corresponding radio band images, are presented in Appendix A (Figs. A.1 to A.21).
We used GALFIT software \citep{Peng2002} to model the light profiles of the ten brightest host galaxies, which have sufficient signal-to-noise ratios. The effective radius ($R_e$) and the Sérsic index ($n$) of the components modeling the light distribution are listed in Table \ref{tabgalfit} and discussed in Sect. 5.1.

\begin{table*}
      \caption[]{Overview of the sample of JSS radio sources.}
      \label{tablist}
      \centering
         \begin{tabular}{cccllllcccc}
            \hline\hline
            \noalign{\smallskip}
            Source& J2000 &Alt. name & id &m& z &Scale&Size&L$_{5GHz}$&Prec./Orb.& Merger\\
                  &       &          &    & &   &\tiny{(pc/mas)}&\tiny{(pc)}&\tiny{(10$^{26}$W/Hz)}&markers&marker\\
            \noalign{\smallskip}
            \hline
            \noalign{\smallskip}
    {\bf 0019--000}&J002225.42+001456.2&4C+00.02& G& 18.4r&0.306&4.55&360&3.6   &CT$_O$&D\\
    {\bf 0108+388}&J011137.32+390628.1&&G& 22.0r&0.669&  7.11         &55&25    &LST&...\\
    0316+162&J031857.80+162832.7&CTA21&G&23.6r&0.907&    7.94          &2800&132   &...&...\\
    {\bf 0428+205}&J043103.76+203734.3&DA138&G&19.3r&0.219&   3.56           &1000&3.0  &LAT&D\\
    {\bf 0500+019}&J050321.20+020304.7&&G&21.0i &0.585&     6.68        &95&26      &S&d\\
    {\bf 0710+439}&J071338.16+434917.2&& G&20.4r&0.518&      6.29         &200&15    &LAT$_O$&D\\
    0941--080&J094336.94--081930.8&&G&17.9r &0.228&           3.67       &350 &1.7     &LAT$_O$&D\\
    1031+567&J103507.04+562846.8&&G&20.2r &0.459&            5.88    & 470 & 9.4    &LT$_O$&D\\
    {\bf 1117+146}&J112027.81+142055.0&4C+14.41&G&20.1R &0.363& 5.10 &500& 4.3       &...&...\\
    {\bf 1245--197}&J124823.90--195918.6&&Q&20.5V &1.275&  8.54        &1500 &212        &AT$_O$&d\\
    1323+321&J132616.51+315409.5 &4C+32.44& G&19.2r&0.369&    5.15    &450&9.9        &LAT&D\\
    {\bf 1345+125}&J134733.36+121724.2 &4C+12.50& G&15.5r&0.122&  2.20&280&1.2       &LS&D\\
    1358+624& J140028.65+621038.6&4C+62.22&G&19.8r &0.429&   5.65    &480&11        &LAT$_O$&D\\
    {\bf 1404+286}&J140700.39+282714.7 &OQ208& G&14.6r&0.077&1.46&12&0.41     &LT$_S$&D\\
    {\bf 1518+047}&J152114.42+043021.7 &4C+04.51& G&22.6r&1.296&  8.56   &1400&140      &...&...\\
    1607+268&J160913.32+264129.0&CTD93&G &20.4r&0.473&        5.98   & 360 & 16      &T$_O$&d\\
    {\bf 2008--068}&J201114.21--064403.5 && G&20.9i&0.547&     6.47   & 200 & 15      &T$_S$&d\\
    {\bf 2128+048}&J213032.88+050217.5 && G&23.3r&0.99&       8.14   & 400 & 93       &LCT$_S$&d\\
    {\bf 2210+016}&J221237.97+015251.3 &4C+01.69&G&21.7r &1.126&8.37 & 800 & 77       &LCT&...\\
    2342+821&J234403.77+822640.4 && Q&20.1r&0.735&            7.39   & 1300 &31        &...&...\\
    {\bf 2352+495}&J235509.46+495008.3&&G &18.4R&0.238&      3.79   & 340 & 2.3      &LSAT$_O$&D\\
            \noalign{\smallskip}
            \hline
         \end{tabular}
\tablefoot{Columns are: source name (radio sources observed with projects BS085 and BO030 are in bold), source name in J2000 coordinates, alternative name, optical identification, optical magnitude, redshift, scale, size, luminosity at 5 GHz in the rest frame, precession/orbital markers from radio images, and merger marker from optical images (see Sect. 4.1). For optical magnitudes, capital letters indicate the Johnson/Cousin photometric system, while lowercase letters indicate the Gunn photometric system.}
   \end{table*}

\begin{table}
      \caption[]{Additional VLBA archive projects.}
      \label{tabproj}
      \centering
         \begin{tabular}{cccc}
            \hline\hline
            \noalign{\smallskip}
            Source&$\nu$  &Project &Epoch\\
                   &\tiny{(GHz)}&  &   \\           
            \noalign{\smallskip}
            \hline
            \noalign{\smallskip}
    0108+388  &15.36&BT030&03 Feb 1997    \\
    \hline
    \noalign{\smallskip}
    0316+162  &0.609&BV014&23 Oct 1995    \\
              &2.29&UF001&12 Aug 2017\\
              &4.98&BG239&14 Feb 2016\\
    \hline
    \noalign{\smallskip}
    0428+205  &0.332&BW067&02 Aug 2003   \\
              &2.29&BG219&23 Jan 2015   \\
    \hline
    \noalign{\smallskip}
    0941--080 &1.41&BM185&26 Apr 2003     \\
              &2.29&UG002&08 Apr 2018     \\
              &8.30&BM185&26 Apr 2003     \\
    \hline 
    \noalign{\smallskip}
    1031+567 &1.41&BM185&26 Apr 2003    \\
             &4.95&BM185&26 Apr 2003    \\
             &  " &BB131&16 Dec 2000    \\
             &8.30&BM185&26 Apr 2003    \\   
     \hline 
    \noalign{\smallskip}        
    1117+146 &2.27&BK095&10 Sep 2002    \\
    \hline 
    \noalign{\smallskip}
    1323+321 &1.67&BM125&08 Jul 2000    \\
             &8.30&BM125&08 Jul 2000   \\  
            &4.54&BM125&08 Jul 2000   \\  
            &15.36&BK068&04 Mar 2001   \\ 
    \hline 
    \noalign{\smallskip}        
    1358+624 & 0.611&BM102&27 Jun 1998 \\
            & 1.41&BM102&27 Jun 1998 \\
             & 4.98&UZ001&14 May 2017         \\
             & 22.22&UZ001&14 May 2017         \\
     \hline 
    \noalign{\smallskip}        
    1404+286 & 1.460&BH104&09 Jan 2003 \\  
    \hline 
    \noalign{\smallskip}
    1607+268 &1.63&BW138&22 Jan 2022   \\
             &2.30&BN025&25 Oct 2003   \\
             &4.99&BN025&25 Oct 2003   \\
             &8.42&BN025&25 Oct 2003   \\
     \hline 
    \noalign{\smallskip}        
    2128+048 &2.30&BF071&31 Jan 2002   \\
    \hline 
    \noalign{\smallskip}
    2342+821 &1.29&GP021&23 Feb 1999   \\
             &2.28&BP222&01 Sep 2018   \\
    \hline 
    \noalign{\smallskip}         
    2352+495 &1.41&BM185&26 Apr 2003 \\
             &2.28&BA064&17 Jan 2003  \\
            \hline
         \end{tabular}
\tablefoot{Columns are: source name, observing frequency, project code, and epoch of observation.}    
   \end{table}

  \begin{figure}
   \centering
   \includegraphics[width=\linewidth]{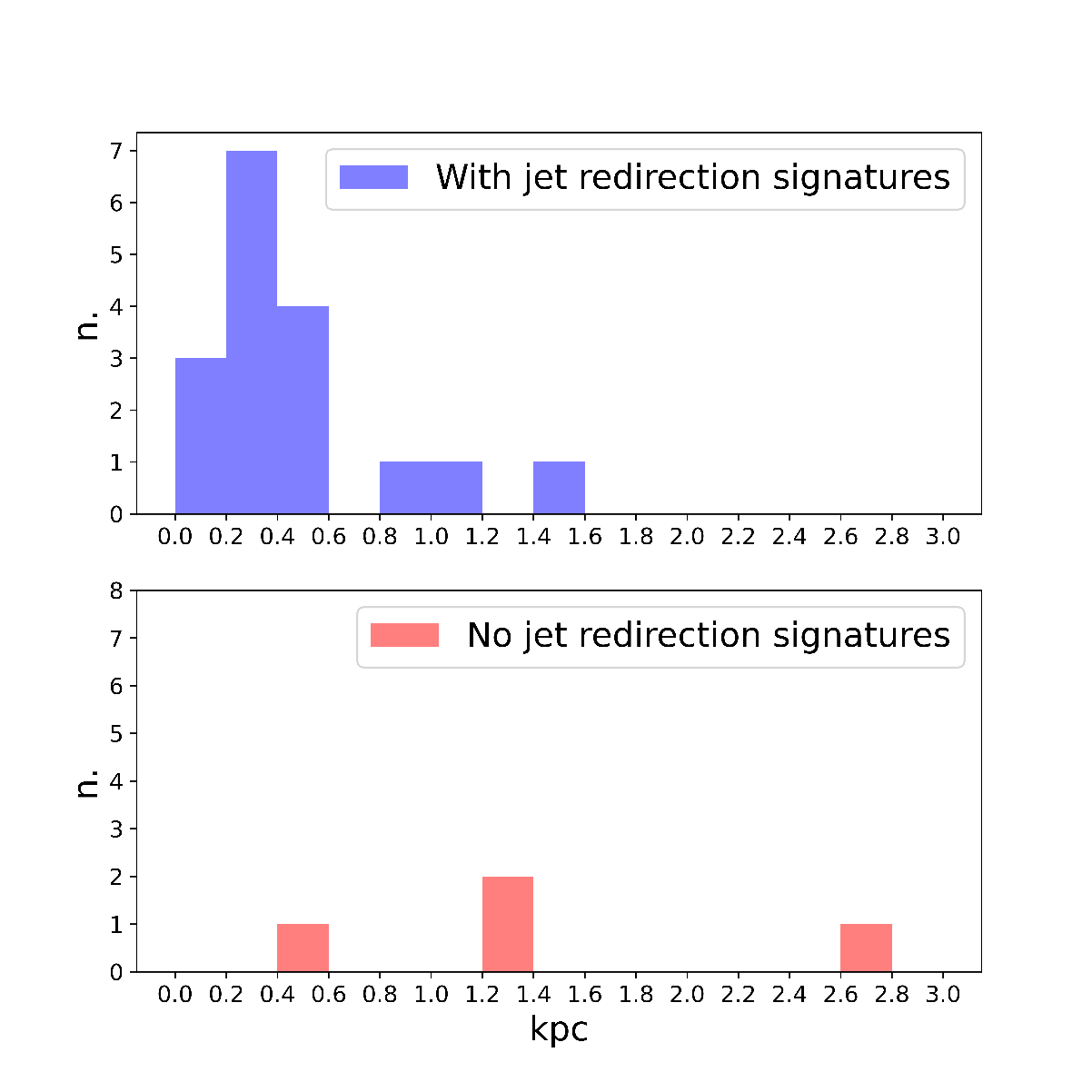}       
   \caption{Projected linear size distribution. The top panel shows the distribution for radio sources with jet redirection signatures. The bottom panel shows the distribution for radio sources with no signatures of jet redirection.}
              \label{linsiz}
    \end{figure}

\begin{table}
      \caption[]{Modeling optical hosts using GALFIT.}
      \label{tabgalfit}
      \centering
         \begin{tabular}{cccccc}
            \hline\hline
            \noalign{\smallskip}
            Source&Band&Type&$R_e$&$n$&Resid.\\
            \noalign{\smallskip}
            \hline
            \noalign{\smallskip}
    0019-000 &r&S\'er.&1.4"&1.1&8\%    \\
        &&"   &3.6"&1.6&    \\ 
             &&"   &1.5"&1.4&    \\       
            \hline
            \noalign{\smallskip}                 
    0428+205 &i&S\'er.&8.6"&6.8&12\%    \\
       &&"   &4"&0.2&    \\ 
             &&point &...&...&    \\  
                         \hline
            \noalign{\smallskip}     
     0710+439     &i&S\'er.&4.3"&7.3&17\%    \\
        &&"   &2.1"&19&    \\ 
                  &&" &1.1&20&    \\  
           \hline       
           \noalign{\smallskip}    
     0941-080 &r&S\'er.&3.1"&0.3&11\%    \\
        &&"   &0.7"&1.5&    \\
         &&"   &0.7"&9.8&    \\  
         \hline
           \noalign{\smallskip}   
     1031+567 &r&S\'er.&1.5"&0.4&7\%    \\
             &&point&...&...&    \\     
     \hline
            \noalign{\smallskip}
     1323+321 &i&S\'er.&2.4"&12&12\%    \\
             &&"&0.6"&1.2&    \\
     \hline
            \noalign{\smallskip}
     1323+321 &i&S\'er.&0.9"&1.4&20\%    \\ 
     \hline
            \noalign{\smallskip}  
    1345+125 &r&S\'er.&0.9"&2.1&9\%    \\
             &&"&0.9"&2&    \\
              &&"&3.9"&0.4&    \\
              &&"&4.1"&0.8&    \\
              \hline
              \noalign{\smallskip}
   1358+624 &r&S\'er.&2.8"&0.04&6\%    \\
             &&"&0.6"&1&    \\  
     \hline
     \noalign{\smallskip}
   1404+286 &i&S\'er.&2.8"&0.9&11\%    \\
             &&"&3.8"&0.2&    \\
              &&"&3.9"&1.3&    \\    
              &&point&...&...&    \\  
              \hline
              \noalign{\smallskip}
 2352+495 &i&S\'er.&11"&11&7\%    \\
             &&"&3.4"&0.1&    \\
              \hline
              \noalign{\smallskip}
 2352+495 &i&S\'er.&8.6"&9&9\%    \\    
            \hline
         \end{tabular}
 \tablefoot{Columns are: source name, optical band, component type (S\'ersic or point-like), effective radius $R_e$ in arcseconds, S\'ersic index $n$, and maximum residual percentage after model subtraction. Both single-component and multi-component models are listed if residuals do not significantly improve. See Sect. 4.2 for details on each object.
 }            
   \end{table}

  \begin{figure}
   \centering
   \includegraphics[width=\linewidth]{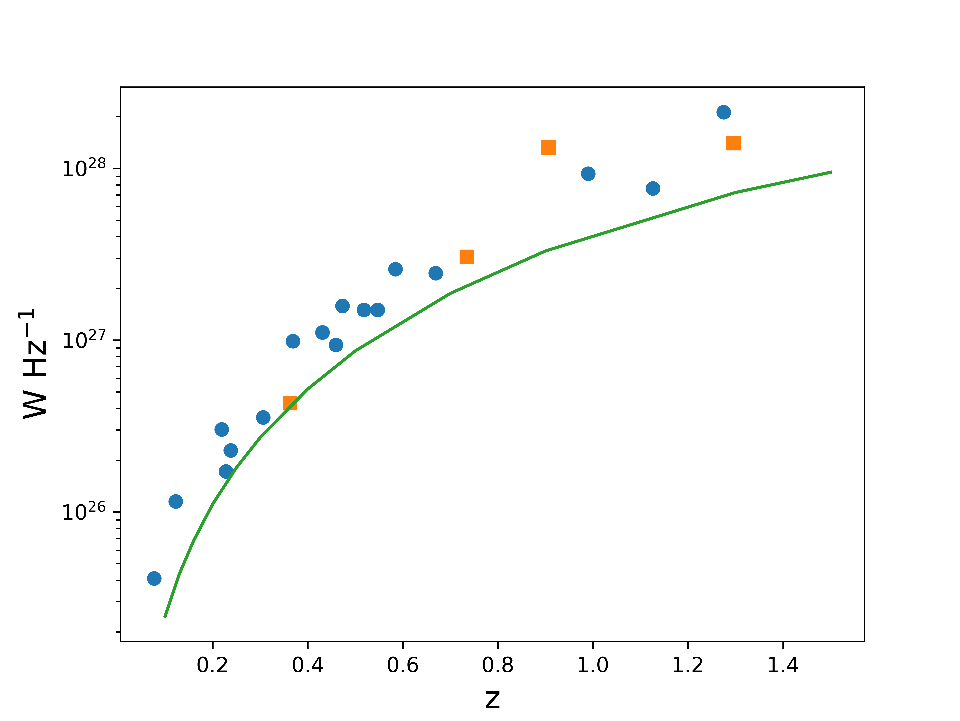}       
   \caption{Luminosity at 5-GHz rest frame. Circles indicate objects with jet redirection signatures, while squares indicate objects with no jet redirection signatures.}
              \label{lum}
    \end{figure}
    
\section{Results}

\subsection{Recurrent patterns in radio morphology}

The JSS radio sources studied in this work exhibit a variety of morphologies. Although each object has its unique characteristics, which we will detail in the notes on individual sources, we recognize several distinctive features. Following the approach in \citet{2019MNRAS.482..240K}, we look for signatures that indicate the possible presence of a precessing radio jet and/or, in our case, the possible occurrence of an orbital motion of the radio AGN.
We consider five morphological markers in the radio (the first two in common with \citealt{2019MNRAS.482..240K}) and one in the optical, as follows:

(C) - Presence of a curved jet.

(S) - Overall appearance characterized by an S-shaped morphology.

(T) - Presence of a significant transverse extension of the radio lobe, i.e. significant low-surface-brightness emission extending from the hot spot to one side perpendicular to the radio source axis. When there is clear evidence that this emission extends in opposite directions with respect to the radio source axis (displaying radial symmetry), we denote this as T$_O$. Conversely, when the emission extends from the hot spot in the same direction with respect to the radio source axis (displaying axial symmetry), we denote it as T$_S$.

(A) - Presence of an arc-like component in the emission around the hot spot or extending from it.

(L) -  A ratio $<$2 between the length and the ``width'' of the radio source, based on the image with the highest sensitivity to extended emission. Precessing jets are expected to make radio emission more extended perpendicular to the jet axis, compared to radio sources with straight jets.  The length is the distance between the hot spots when these are present, or the maximum extension of radio emission, while the width is the maximum extension of the radio emission perpendicular to the source axis.

(D) - Disturbed morphology of the host galaxy. When the optical host is faint but still shows hints of an irregular shape or disturbed morphology, we denote it with a lowercase {\it d}.

The last two columns of Table \ref{tablist} indicate the presence of these morphological markers in our sample, with 17 objects showing two or more of these signatures and four objects showing none.

\subsection{Notes on individual sources}

In the following, for each object, we provide relevant information from the literature and a description of our findings. In images, tables, and text, isolated regions and components are labeled with letter designations. Unless otherwise indicated, the basic information reported is drawn from the NASA/IPAC Extragalactic Database (NED)\footnote{ned.ipac.caltech.edu}. 

{\bf 0019-000}: There is little information in the literature on the parsec-scale morphology of this source. \cite{2002AAS...201.7611F} present images at 2.3 and 8.7 GHz, showing a structure resembling a core-jet morphology. However, our VLBA images at L,C,X radio bands reveal a more complex morphology. The northern component is extended and has a steep spectrum, indicating that it is unlikely to be the core. We identify the two brightest components at the edges of the radio structure as the hot spots. A curved, interrupted trail of emission connects these brighter components, with additional emission extending transversely to the jet axis in L band. The most likely explanation is that projection effects distort a double morphology.  
There is no evidence of a compact and flat spectrum component that could be identified as the core.
The optical host appears as a double-interacting galaxy with a few much weaker objects within a distance of 20 arcseconds. The light profile of the host galaxy is best modeled with two S\'ersic profiles for the brighter component and one for the weaker component (Table \ref{tabgalfit}).

{\bf 0108+388}: 
The radio morphology comprises two curved and knotty jets that terminate at two hot spots. At 5 and 8.3 GHz, extended emission widening toward the northwest is detected. \citet{1996ApJ...463...95T} identify an inverted spectrum component at the center of the radio emission as the core. However, this component is not clearly distinguishable in our images, possibly due to variability. \citet{2003PASA...20...69P} report a separation speed of 0.26$c$ between the two hot spots.
The optical host is a faint galaxy, with a similarly weak object located 20 arcseconds away. Notably, the arcsecond-scale radio emission previously detected near the compact radio source \citep{1990A&A...232...19B} coincides with this optical object, challenging its direct association with the compact radio source, as also noted by \citet{2017MNRAS.471.3806K}. Other unresolved bright objects in the optical field are likely foreground stars.

{\bf 0316+162}: The radio morphology exhibits an asymmetric double radio source, characterized by high flux density and arm-length ratios between the two sides. The images reveal a faint, rather amorphous southern part, while the northern part resembles a jet termination. 
\citet{2021MNRAS.504.2312D} present a global VLBI image at 327 MHz, which is consistent with our 609 MHz image.
\citet{2013MNRAS.433..147D} identified the Ce3 component in our images as the core. However, \citet{2021RNAAS...5...60F} revised this identification, suggesting that the Ce1 component that we detect at 5 GHz, but not at lower frequencies, might be the core. The lack of detection below 5 GHz in our images indicates variability or a highly inverted spectrum, consistent with their findings.
The optical host is a faint galaxy surrounded by other objects within a 10 arcsecond distance.  

{\bf 0428+205}: Based on high-frequency radio images or considering only the brightest components at lower frequencies, this object would resemble the previous one (0316+162) with two sides showing very different flux densities. Indeed, our reprocessed archive image at 1.2 GHz reveals an arc of faint emission extending eastward from the northern hot spot/lobe and an emission extending westward from the southern hot spot, following an arc before turning east.  This is corroborated by the 332 MHz image, where the lateral emission is much brighter. The 1.7 GHz image presented by \citet{1995A&A...295...27D} already shows some hints of the faint lateral emission.
The high-frequency images show the southern side in detail with a jet-like morphology terminating in a hot spot.
\citet{2013MNRAS.433..147D}  proposed component Ce1 as the source core based on the flat spectral index found between their 1.7 and 5 GHz VLBI data.
The host galaxy exhibits optical emission extending from the center toward the southeast and then northeast, enclosed within a roughly triangular-shaped envelope. Some fainter objects are located within a 10-arcsecond distance. The light profile of the host galaxy is best modeled with two S\'ersic profiles and one unresolved component, which is possibly associated with the presence of an optical AGN (Table \ref{tabgalfit}).

{\bf 0500+019}: Our VLBA image at 8.42 GHz clearly shows an S-shaped emission, which confirms the morphology presented by \citet{1997A&A...325..943S,2001A&A...377..377S}. At higher frequencies, the faint southern part is gradually resolved out.
\citet{2012A&A...544A..34P} present an image at 2.3 GHz showing fainter emission extending west on the northern side, and extending east on the southern side. 
The optical host is a slightly elongated object surrounded by many other objects within a distance of 20 arcseconds. 

{\bf 0710+439}: The radio morphology consists of two hot spots and a central jet. \citet{1996ApJ...463...95T} propose the southern tip of the central jet-like feature as the core. The images at low frequency show faint extended emission following an arc toward the west from the northern hot spot and toward the east from the southern hot spot. A hint of this lateral emission is already seen in an image presented by \citet{1996ApJS..105..299F}.
 An expansion speed of the hot spots of 0.43$c$ is detected by \citet{2003PASA...20...69P}. 
The optical host has an irregular, roughly triangular shape.
Several distinct components are present within the same galactic envelope or in close proximity to it. The model listed in Table \ref{tabgalfit} includes only the three S\'ersic components located within the common galactic envelope.

{\bf 0941-080}: The radio morphology consists of two main components with an arc shape. The emission extending from the southern component toward the center of activity can be separated in an additional compact component in the 8.3 GHz image.
The optical host is a galaxy with a double nucleus, with the radio AGN associated with the northern one. 
The light profile of the host galaxy is best modeled with three S\'ersic profiles (Table \ref{tabgalfit}).

{\bf 1031+567}: 
The radio morphology consists of two bright hot spots, a compact central component, and significantly extended emission that elongates laterally and in opposite directions relative to the two hot spots. The size of the lateral emission is comparable to the distance between the hot spots.  A hint of this emission was already observed in an image at 2.3 GHz presented by \cite{2011A&A...535A..24S}.
\cite{1996ApJ...463...95T} identified component C as the possible core, although doubts arise due to its steep spectral index, which is also confirmed by our estimate.
\citet{2003PASA...20...69P} report a relative proper motion of the two hot spots of 0.27$c$.
This is a striking example of a source with an aligned morphology seen at high frequency, where only the brightest and recently active regions are visible, but showing much more complexity when sensitive observations at low frequency allow the detection of fainter emission. 
The optical host is a faint, slightly resolved object with a couple of other objects at a distance of 15 arcseconds. The closest unresolved bright object is likely a foreground star.  Modeling the light profile of the host galaxy requires a point-like component, potentially associated with an optical AGN contribution, in addition to a S\'ersic component  (Table \ref{tabgalfit}). 

{\bf 1117+146}: The radio morphology shows a double structure with two bright hot spots at the outer edges and jet/lobes extending toward the center of activity.
\citet{1995A&A...295..629S} presented a global VLBI image at 1.6 GHz that is fully consistent with our images. \cite{1998MNRAS.297..559B} presented a MERLIN image at 23 GHz, showing a central component that is not seen at lower frequencies, which they identify as the core.
Despite the high sensitivity of our 5 GHz image, we do not detect that compact component, implying that it is heavily self-absorbed and/or significantly variable in flux density, confirming its identification with the radio core. 
This radio source is the only one of our sample that shows a clear scaled-down FR II morphology.
The optical host is a galaxy with a couple of fainter objects within 10 arcseconds.

{\bf 1245-197}: \cite{2011A&A...535A..24S} presented images of this radio source at 2.3 and 8.4 GHz, showing a morphology similar to a double/triple radio source.
Our high-sensitivity images reveal a different morphology. The overall structure, seen in our 1.7 GHz image, consists of two separated regions of radio emission aligned in the E-W direction, 150 mas apart, with the eastern one being six times brighter. At higher frequencies, the western side is partially resolved out, while the eastern part displays an arc-like substructure with a brighter component at one end and a trail of emission extending north before bending east. 
Despite the identification of the optical host with a quasar, the PanSTARRS image shows a faint object with a rather triangular shape. Other weakly resolved sources, likely galaxies, are present within a distance of 15 arcseconds.

{\bf 1323+321}: The radio morphology consists of two hot spots prominent in the images at the highest frequencies, with jet-like structures extending back toward the center of activity.  At lower frequencies, we observe extended emission broadening transversely to the hot spot positions, reaching a size almost comparable to the distance between the two hot spots. This extended lateral emission is already visible in an image at 2.3 GHz presented by \cite{1996ApJS..105..299F}.
\cite{2016MNRAS.459..820T} identify a flat spectrum core coincident with component C in our images. The spectral index of the core calculated from our images is not reliable due to its faintness.
\cite{2004ApJ...609..539K} find no evidence of relative proper motion of the two hot spots.
The optical host is a galaxy surrounded by small objects within a distance of $\sim$10 arcseconds, with a diffuse trail extending from south to east and then to north. 
The light profile of the host galaxy is best modeled with two S\'ersic profiles. Despite the presence of faint extended emission, an acceptable fit can also be achieved with a single S\'ersic profile characterized by a relatively low S\'ersic index (Table \ref{tabgalfit}).

{\bf 1345+125}: 
The radio morphology consists of a bright, heavily bent southern jet terminating in a hot spot and surrounded by a lobe. A much fainter northern lobe is also present. The radio core is associated with component C \citep{2001A&A...377..377S,2002A&A...385..768X,2003ApJ...584..135L}.  \cite{2003ApJ...584..135L} estimate a jet speed close to the core of 0.84$c$ and a viewing angle of 64$^\circ$. 
\citet{2013Sci...341.1082M} find a fast outflow of cold gas (HI) associated with the southern hot spot.
\citet{2005A&A...443..891S} report the detection of faint emission at the arcsecond scale, considered as a relic of the past radio activity of the AGN. 
The optical host is an ultraluminous infrared galaxy showing a double nucleus, an irregular shape, trails, and tails, which are clear signatures of an ongoing merger.  \cite{1993ApJS...88....1S} associate the radio source with the western nucleus, a conclusion confirmed by \cite{2000AJ....120.2284A} using HST observations. \cite{2000AJ....119..991S} discuss infrared observations, finding that the western nucleus is unresolved and significantly redder than the eastern nucleus, which appears extended.
\cite{2016A&A...596A..19E} present detailed optical images and conclude that a major, possibly multiple, merger event is ongoing.
Given the complex optical morphology of this interacting system, several S\'ersic profiles are required to model the host galaxy. Table \ref{tabgalfit} lists the four S\'ersic profiles that account for the brighter part of the optical emission. 

{\bf 1358+624}:  
The radio morphology consists of a knotty and bright jet on the southeast side terminating in a hot spot, with fainter extended emission further along the jet direction and south of the hot spot. The northwest side has a hot spot with additional emission that extends northward, broadens, and then bends to the east. 
The compact component C has an inverted spectrum and is identified as the core, which confirms earlier works by \citet{1996ApJ...463...95T} and \citet{2013MNRAS.433..147D}.
The host galaxy has a boxy morphology with a hint of a transverse band of optical absorption. Another object with a slightly elongated morphology is located 5 arcseconds to the north.  
The light profile of the host galaxy is best modeled with two S\'ersic profiles (Table \ref{tabgalfit}).

{\bf 1404+286}:  The radio morphology consists of a brighter eastern component and a much fainter western component. The eastern component exhibits a substructure with transverse emission relative to the radio source axis and additional emission extending toward the center. The western side consists of two similar components. We identify the brightest eastern and western components as the two hot spots of a double source. There is a hint of emission between the two hot spots. \citet{2013A&A...550A.113W} also report faint components between the hot spots, not always visible at different epochs, and a separation speed of 0.134$c$ for the two hot spots, along with evidence of sideways motions. Very faint emission is detected $\sim 30$ mas west of the main structure at 1.7 GHz. The size of the radio source reported in Table \ref{tablist} does not include this faint emission.
The optical host is identified as a broad-line radio galaxy with a disk morphology and a trail of emission bending from north to east. There are two additional components within the common envelope; the eastern one is unresolved and possibly associated with a foreground star.  
The light profile of the host galaxy, excluding the unresolved object east of the optical core, is best modeled using three S\'ersic profiles along with an additional unresolved component, which may be linked to the presence of an optical AGN (Table \ref{tabgalfit}).

{\bf 1518+047}: The radio morphology consists of two bright hot spots, with additional emission extending outward from the northern hot spot and toward the center from the southern hot spot. 
\cite{2010MNRAS.402.1892O} report no evidence of new electron injection in the hot spots, suggesting that this object is a dying radio source.
The optical host is a faint galaxy surrounded by objects of similar magnitude within 10 arcseconds.

{\bf 1607+268}: The radio morphology consists of a double structure, with diffuse emission extending southward from the northern hot spot, then bending to the west. The southern side shows a jet terminating in a hot spot with additional emission elongating laterally toward the east.
The optical host is a faint and roughly triangular object with a couple of objects within 10 arcseconds. 

{\bf 2008-068}: The radio morphology consists of a brighter northern hot spot and a fainter southern one. Additional radio emission extends from the two hot spots at an angle of approximately $45^\circ$ relative to the radio source axis. A compact component is present at the center of the structure, with a flatter spectral index compared to the other components, but it is still too steep to confidently identify it as the core. 
The optical host is a faint object with an irregular double structure. 

{\bf 2128+048}: The radio morphology is similar to that of the previous object, and consists of a brighter northern hot spot and a fainter southern one. Again, emission extends from the two hot spots at an angle of approximately $45^\circ$ relative to the radio source axis and then bends toward the center of the radio source. There are also a couple of compact components between the hot spots, none of which has a flat spectrum, and a bent jet connecting the northern compact component to the northern hot spot. The radio morphology of this source was investigated by \cite{1997A&A...325..943S} at 5 GHz, showing a structure similar to our images, and by \cite{1998A&AS..129..219D}, where the strongest components are visible at a lower angular resolution.
The optical host has a double morphology with two weak components at a distance of 5 arcseconds, with the radio source coincident with the peak of the northern optical emission.

{\bf 2210+016}: The radio morphology is quite complex. A compact yet resolved component, probably a hot spot, is located at the western edge of the radio source. A jet extends eastward from the center of the radio source and then broadens toward the southeast. 
The optical host is a very faint object.  

{\bf 2342+821}: The radio structure consists of three main components. The westernmost component features a bright hot spot surrounded by lobe emission. Each of the other two fainter components, located at the center and the eastern edge, exhibits a double substructure. The image at 327 MHz presented by \citet{2021MNRAS.504.2312D} shows additional extended emission between components E and Ce.
The optical host is identified as a quasar. Some faint objects are present within 10 arcseconds.  

{\bf 2352+495}: 
The radio morphology consists of two bright hot spots. Extended emission laterally elongates from the southern hot spot, while it follows an arc from the northern hot spot. From the bright central component, there is emission extending northward, and a long curved jet connects the center to the southern hot spot.
\citet{1996ApJ...463...95T} identify component C in our images as the core, which we confirm based on its compactness and flat spectral index. \citet{2003PASA...20...69P} estimate a separation speed of 0.17$c$ for the two hot spots.
The central region observed by \cite{2000ApJ...541..112T} shows a complex structure in which five components have velocities ranging from 0.27$c$ to 0.76$c$. 
The optical host is a galaxy with an irregular, somewhat triangular shape with hints of diffuse emission around the main structure and several objects within 10 arcseconds.  
The light profile of the host galaxy is best modeled using two S\'ersic profiles. 
A satisfactory fit, with slightly higher residuals, can also be achieved using a single S\'ersic profile with a high S\'ersic index (Table \ref{tabgalfit}).

\section{Discussion}

\subsection{Radio and optical morphology}

Seventeen of the 21 objects in our sample exhibit signatures consistent with jet precession or relative motion between the AGN and the ISM, while four show no such evidence. Therefore, less than 20\% show an aligned morphology. This fraction might be even lower, as these objects could be reclassified if additional extended emission is detected at lower frequencies or with more sensitive observations, including better UV-coverage, as was the case for 0428+205 and 0710+439. 
The presence of an S-shaped morphology in CSOs has previously been discussed by \citet{1996ApJ...463...95T} as indicative of ongoing jet precession. This phenomenon was studied in more detail in 2352+495 by \citet{1996ApJ...460..612R} who, however, favored a young age for this radio source. 
In our sample, only half a dozen objects show either a core or a core candidate, confirming that flat-spectrum cores are mostly undetected in JSS radio sources at the frequencies and sensitivities of the currently available observations. 
 
Despite the limited statistics, we observe a clear difference in the (projected) linear size between sources with and without indications of jet redirection. The former are more compact, usually < 1 kpc (Fig. \ref{linsiz}), with no evident difference in the luminosity distribution, as shown in Fig. \ref{lum}. The 5 GHz luminosity presented in Table \ref{tablist} and Fig. \ref{lum} was calculated using the total 5 GHz flux density and the high-frequency spectral index reported by \citet{1998A&AS..131..303S}.

Among the 17 objects exhibiting signatures of jet precession, ten have host galaxies observed with S/N $>$ 50, allowing for a detailed examination of their features.
None of these ten galaxies is a typical elliptical galaxy. All of them exhibit disturbed optical morphologies and have nearby fainter objects within a few arcseconds. Five of these galaxies show double nuclei or multiple compact components within the same optical envelope. Others show varying degrees of morphological disturbances (tails, arms, boxy/triangular shapes), indicative of current or recent mergers (e.g., \citealt{2016MNRAS.456.3032P}).
In particular, the light profiles of these ten brightest host galaxies are best modeled using multiple S\'ersic components (Table \ref{tabgalfit}). Even in the few cases where a single Sérsic component provides an acceptable fit, the S\'ersic index significantly deviates from 4 (corresponding to a de Vaucouleurs profile, typically indicative of a passive elliptical galaxy or a fully relaxed merger). It should be noted that none of the components exhibits a Sérsic index close to 4.
The presence of two or more light profiles with (i) intermediate S\'ersic indices and (ii) different effective radii -- one confined to the brightest central region and the other extending to the outer parts of the host -- is consistent with predictions from hydrodynamical simulations of merger remnants \citep{Hopkins2008, Hopkins2009}. 

Indeed, it is well established that GPS radio galaxies reside in optically disturbed or merging systems \citep[and references therein]{2021A&ARv..29....3O}.
In a relevant recent study, \citet{2024A&A...682A..18M} found a significantly higher fraction of mergers ($\sim60\%$) in their sample of PSS-CSOs (a large sample including CSOs and Peaked Spectrum Sources, the latter comprising HFP, GPS, and CSS sources) compared to their control sample of large radio galaxies ($\sim15\%$).
Our results are consistent with their findings, but deeper and multi-band observations are required for full confirmation. Based on the fractions of mergers in their samples, \citet{2024A&A...682A..18M} also concluded that only half of the peaked sources are expected to evolve into large radio galaxies.

\subsection{A novel interpretation}

The morphology observed in most JSS radio sources is naturally explained by a jet precessing rapidly at a large angle (i.e. $>10^\circ$), with the faint emission trail transverse to the hot spot indicating the past position of the hot spot. 
The presence of arc segments, clearly visible in some objects (i.e. 0941-080, 1245-197), further supports the hypothesis of a precessing jet.
In the case of radio sources with transverse lobes that extend on the same side (T$_S$ marker), the dominant effect might be the motion of the black hole through the ISM. 

\begin{figure*}
   \centering
   \includegraphics[width=0.85\linewidth]{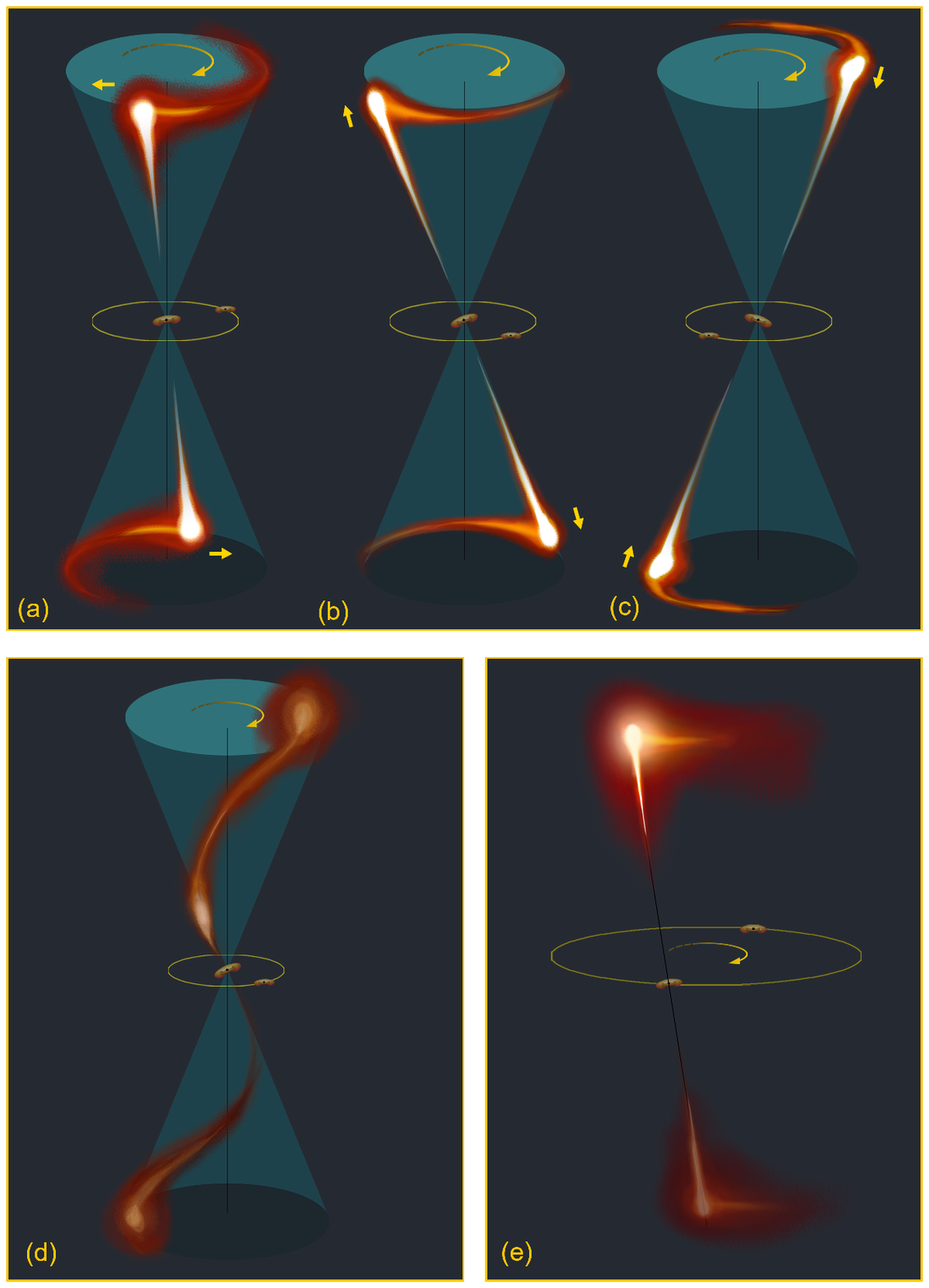}       
\caption{Jet redirection scenario. Top row: Jet precession alone determines the observed morphology, with transverse (a), expansion (b), or contraction (c) motions on the plane of the sky resulting from geometric and projection effects. Bottom row: Panel (d) represents the case of a precessing jet with a period shorter than (or comparable to) the travel time of the jet from the core to the hot spot. In panel (e) the orbital motion of the radio-loud AGN alone determines the observed morphology (see Sect. 5.2 for details). The image is not to scale and is for illustrative purposes only [Illustration by Luca Jerrert Rossi]. }
              \label{draw}
    \end{figure*}

Based on the radio and optical morphologies described above, we propose a scenario in which radio sources originating in galaxies undergoing a merging phase may remain confined (frustrated) within the galaxy core due to jet redirection driven by a combination of jet precession and the orbital motion of the AGN producing the jet. Our scenario is illustrated in Fig. \ref{draw}, in the idealized case of jet-precession-driven morphology only (Fig. 3a, 3b, 3c, and 3d) or orbital-motion-driven morphology only (Fig. 3e). In actual cases, both mechanisms may contribute, with one potentially dominating over the other.

In a merging system, large regions of gas might be in relative motion with respect to the general gravitational potential well, contributing to the relative motion between the hot spots and the environment. The dynamically active gas and its inhomogeneous distribution in a merging system can lead to complex radio morphologies, characterized by significant differences in size, shape, and brightness between the two opposite sides of the radio source relative to their central origin.

The morphology seen in our images suggests that the hot spot, in its lateral motion, traverses a path comparable to the size of the radio source. Consequently, a jet that continuously changes the direction and location of its terminal shock cannot efficiently push the gas forward and expand rapidly outward. 
In the following, we will critically discuss our scenario based on observational evidence.

\subsection{Proper and transverse motions}

Several objects have measured proper motions of the hot spots expanding outwards
\citep{2003PASA...20...69P,2005ApJ...622..136G,2012ApJ...760...77A}, suggesting that contrary to our hypothesis, at least for those sources, the youth scenario is the most likely explanation. 
\cite{2021A&ARv..29....3O} list 42 objects with measured proper motions, including nine with only an upper limit (excluding proper motions of inner knots). Six of these objects are part of our sample: 0108+388, 0710+439, 1031+567, 1404+286, 1607+268, and 2352+495.
Regarding 1404+286, \citet{2013A&A...550A.113W} report a complex relative motion of the hot spots, including a transverse component. Similarly, \citet{2003PASA...20...69P} report an unusual side motion for 0108+388 and 1031+567, which they attribute to the ``dentist's drill effect'' \citep{1982IAUS...97..163S}. \cite{2020IAUS..342..237C} tentatively finds an oscillation in the distance between the two hot spots over time for 1031+567. Furthermore, a revised analysis of the proper motion of 2352+495 suggests a transverse relative displacement of the hot spots (Stanghellini et al. in prep.). It appears that most objects with detected proper motions may need a revision focused on a more precise measurement of the lateral displacements of the jet terminations.

Transverse motions, or more precisely, transverse displacements of the termination of the jet (in our precessing scenario, the hot spot does not actually have dynamic motion), have also been detected in some JSS radio sources not included in our sample.
\citet{2009AN....330..153S} report transverse displacement for three JSS radio sources and suggest jet precession as the cause. \citet{2012ApJS..198....5A} confirm the transverse displacement in one of the objects mentioned above and find more evidence in two additional cases. 

In our scenario, when jet precession is the dominant cause of the changing direction of the jet, we may expect displacement vector components both transverse to and along the jet axis due to projection effects (Fig. 3a,b,c). Therefore, the proper motions reported in literature, even when confirmed, are not inconsistent with our scenario (Fig. 3b). Indeed, we should expect some sources to appear to be shrinking (Fig. 3c), and some examples have been reported.  \citet{2012ApJS..198....5A} find a decreasing distance with time in one object they observed (J0017+5312), although the classification of this object as a JSS radio source is uncertain. Another intriguing example of a shrinking radio source is PKS 1155+251 \citep{2008ApJ...684..153T,2017MNRAS.471.1873Y}. However, in this case, the contraction cannot be explained solely by precession and it likely requires, within our scenario, a combination of precession and the orbital motion of the AGN in a dual black hole system.
A larger number of JSS radio sources with measured proper motions/displacements are needed to conduct a statistically robust study and place solid constraints on our hypothesis.

\subsection{Jet precession at different scales}
Lobe asymmetries, trails in different directions, curved jets, and S-shaped jets are often seen in large radio galaxies, with the most likely explanation being a precessing jet.
\citet{2019MNRAS.482..240K} found strong evidence for jet precession in three-quarters of their studied sample of FR II radio galaxies. Therefore, the presence of a precessing jet alone is not sufficient to keep the radio source small.

Precessing jets in large radio galaxies have recently been modeled by \citet{2023ApJ...948...25N}, who found that different precession parameters significantly change the size of a radio galaxy at a given time. In particular, jets precessing at a large angle can significantly slow down the overall growth of the radio source.
Thus, the precession angle is an important parameter in determining the growth and expansion of the radio source, with precession at large angles keeping the radio source more compact.

The precession period may be even more important in influencing the size of the radio source.  
3D hydrodynamical simulations by \citet{2020MNRAS.499.5765H} show that a shorter precession period causes the radio source to grow at a slower rate. The arc features and lobe extensions seen in the simulated images (Fig. 2 in \citealt{2020MNRAS.499.5765H}) are similar to the arc features and transverse lobes observed in our images.

\citet{2019MNRAS.482..240K} estimate that the jet precession period in their sample of 3CR radio galaxies is between $10^6-10^7$ yr. 
The best case of measured transverse motion in JSS radio sources is J1335+5844, detected by \citet{2009AN....330..153S} and confirmed by \citet{2012ApJS..198....5A}. The transverse displacement estimated using the values in Table 1 in \citet{2012ApJS..198....5A} corresponds to a projected angular velocity for the jet axis of 4.2($\pm 1.2)$ arcminutes per year (0.008 mas/yr for the hot spot displacement, assuming symmetrical behavior of the two hot spots), resulting in a period of $\lesssim 2\times 10^3$yr, assuming a cone of 45$^\circ$ for the precession. 
Similar precessing periods may be estimated with even less accuracy for the other few objects showing transverse motions, e.g. J0706+4647 and J1823+7938 in \citet{2009AN....330..153S}.
With the caveat of high uncertainty and low statistics, it seems that the precession period of the jet axis in a JSS radio source is $10^{2-3}$ times shorter than in FR II radio sources. 

We investigate whether such a short period is possible.
\citet{2019MNRAS.482..240K} discussed the possible causes of precession. Geodetic precession predicts a precession period much longer than what is consistent with our scenario. Geodetic precession would require an unrealistic distance between the black holes to allow for a 10$^4$ year period, as it would place the binary supermassive black hole (SMBH) in the gravitational-wave regime, leading to fast coalescence. On the other hand, precession induced by an irregular or warped accretion disk would last for only one period and quickly disappear.
\citet{2005A&A...431..831L} predict a precession period of a few 10$^3$ years in the case of 3C345, with the precession caused by the orbital motion of the secondary on the accretion disk of the active black hole. Therefore, there is at least one physical mechanism capable of explaining the precession period that we see in JSS radio sources. However, determining the actual cause of the jet precession is beyond the scope of this paper.

In our scenario, radio sources may be kept compact when the precession angle is large and the precession period is short.
The comparison between the time required for the relativistic plasma to travel from the core to the hot spot and the precession period determines the observed morphology of the jets. If the travel time is much shorter than the precession period, the jets appear straight; otherwise, they will appear S-shaped
\citep{2019MNRAS.482..240K}.

\subsection{Binary supermassive black holes and their orbital motion}

As an intermediate phase of the merging process, pairs of SMBHs are expected to form. Following the nomenclature of \citet{2019NewAR..8601525D}, we refer to systems where the two active SMBHs are separated by more than 1 pc as dual AGN, while those with separations $<1$ pc are classified as binary AGN systems.

Following a galaxy merger, the orbital periods of surviving binary black holes are in the range 10-10$^5$yr \citep{2002MNRAS.331..935Y}. Interestingly, this range aligns with the estimated radiative age of electrons in the lobes of JSS radio sources (e.g. \citealt{2003PASA...20...19M}), which corresponds to the age of the oldest observed radio emission, although the radio source itself may be older. 

\citet{2020ApJ...897...86C} find that the coalescence timescales of binary black holes are in the range $10^8$-10$^{12}$yr.  This implies that if a radio source cannot expand due to the presence of a dual/binary black hole, it may remain compact for a duration comparable to the lifetime of a large extragalactic radio source, or even indefinitely. This potentially explains the relatively high number of JSS radio sources.

The orbital motion of a dual or binary black hole may be another factor contributing to the small size of the radio source, as the relative motion of the black hole and the ISM may lead the terminal shock of the jet to continuously shift the impact direction. A morphology showing transverse lobes or trails on the same side of the radio source axis may be the case where this phenomenon dominates (Fig. 3e). Such a phenomenon, observed on a smaller scale, is similar to the narrow-angle tail (NAT) radio galaxies seen in galaxy clusters. The long tails, a characteristic of NATs, are absent in this case because of the shorter path length and faster radiative losses caused by the stronger magnetic fields (see also Sect. 5.6). 

\citet{2017ApJ...843...14B} calculated the orbital period of the dual black hole system in the CSO 0402+379 to be $3 \times 10^4$ years.  Remarkably, this orbital motion period is of the same order, or within an order of magnitude, as the precession periods in JSS radio sources discussed above. Therefore, it is not surprising that they may produce similar effects on the morphology. 

\citet{2014Natur.511...57D} confirmed that J1502+1115 is a triple SMBH system with a distance of 140 pc for the closest pair, which is 20 times the separation between black holes in the system studied by \citet{2017ApJ...843...14B}. They also revealed an S-shaped radio emission centered on that pair, extending for $\sim$10 kpc, and attributed to jet precession. This size is larger than that of the objects in our sample, but still within the host galaxy. The latter case supports the connection between dual black holes and radio compactness, thus justifying the extension of our study to larger JSS radio sources.

\subsection{Radiative ages and old radio emission}

\citet{2003PASA...20...19M} estimated the radiative age of the radio source 1323+321. Although the purpose of that study is the comparison between radiative and dynamical ages under the assumption that the radio source expands along the jet axes, it is noteworthy that the figures and plots (Fig. 4 in \citealt{2003PASA...20...19M}) show that the radiative age also increases along the transverse trail of the southern hot spot. This could be due to aging of the backflow from the hot spot, as seen in many extended radio galaxies, but it could also indicate that the hot spot was previously located there. Considering a displacement of the hot spot of 10 mas in 2000 years, the jet would be precessing with a projected angular velocity of 0.3 arcminute/yr, consistent with the few estimated angular velocities (Sect. 5.4). 

As to the other objects in our sample, we lack sufficient and/or reliable multifrequency data with matched UV-coverage. This matched UV-coverage is necessary to achieve comparable angular resolution and sensitivity to extended emission, which would allow us to estimate the radiative age along the trails/lobes.

If a radio source spends a long time in the central part of the host galaxy with its precessing jet, there should be residual emission forming a halo or cocoon around the active regions. With magnetic fields of a few to several tens of mG \citep{2016AN....337....9O}, the break frequency would shift below 100 MHz in a few thousands years, making the halo generally undetectable in much older radio sources at typical observing frequencies. Occasionally, this diffuse emission can be observed, as in the case of the lowest redshift object in our sample, 1404+286, where faint emission at mas scale beyond the double structure has been reported and confirmed \citep{2013A&A...550A.113W}. Recent EVN+eMERLIN observations have detected even more diffuse emission around the double structure, consistent with our scenario (Stanghellini et al. in prep.).

\section{Conclusions}
We performed a morphological study of a sample of JSS radio sources, and found that the radio structure often shows signatures of precessing jets at large angles or in motion with respect to the ISM, and the optical hosts often appear as merging systems.
We suggest that JSS radio galaxies consist of a mixed population, with a minority of aligned sources likely to evolve into FR I/II radio galaxies. However, most exhibit complex morphologies and may host a rapidly precessing jet and/or an AGN moving relative to the ISM, either due to orbital motion in a binary system or the dynamic motion of gas during a merger.
These effects may keep them confined within the inner regions of their host galaxies, making them ``frustrated'' radio sources.

We propose a scenario in which most JSS radio sources reside in merging systems hosting dual or binary black holes and remain at parsec scales for an extended time due to continuous jet redirection. Asymmetries in size and flux density may arise from a combination of factors, including moderate relativistic beaming, an inhomogeneous ISM distribution, the orbital motion of the active SMBH, and the relative motion of the hot spots with respect to the ISM.

The possibility that a fraction of JSS radio sources are short-lived objects is not inconsistent with our scenario. In a galaxy merger, the environment surrounding the SMBH may be dynamically heavily disturbed, and the reservoir of accreting material (gas or stars) may vary greatly over time, with short episodes of activity and the radio source possibly switching on and off.

We presented several pieces of evidence supporting the hypothesis that a significant fraction, potentially even the majority, of radio sources confined to the inner regions of their host galaxy may remain in this phase for an extended time due to the rapid redirection of their jets. Further observations, refined analysis, and specific modeling, including numerical simulations, are required to confirm that our scenario is both valid and physically consistent. On the other hand, the evidence collected so far makes this scenario worthy of further investigation, which is currently underway.
If our scenario is correct, we expect to find more examples of transverse motions and more evidence of old emissions tracing the path of the precessing jet. JSS radio sources may be the objects to focus on and look for in the hunt for binary black holes.

\begin{acknowledgements}
We thank Luca Jerrert Rossi (jerrert@gmail.com) for creating the drawing used in this paper. 
CO and SB acknowledge support from the Natural Sciences and Engineering Research Council (NSERC) of Canada.
CSp acknowledges financial support from the Italian National Institute for Astrophysics (INAF) under the project “Collaborative research on VLBI as an ultimate test to $\Lambda$CDM model” (Ricerca Fondamentale 2022).
MPT acknowledges financial support from the Severo Ochoa grant CEX2021-001131-S and from the National grant PID2020-117404GB-C21, funded by MCIU/AEI/ 10.13039/501100011033.
The National Radio Astronomy Observatory is a facility of the National Science Foundation operated under cooperative agreement by Associated Universities, Inc.
This research has made use of the NASA/IPAC Extragalactic Database, which is funded by the National Aeronautics and Space Administration and operated by the California Institute of Technology.
Based on observations made with the Nordic Optical Telescope, owned in collaboration by the University of Turku and Aarhus University, and operated jointly by Aarhus University, the University of Turku and the University of Oslo, representing Denmark, Finland and Norway, the University of Iceland and Stockholm University at the Observatorio del Roque de los Muchachos, La Palma, Spain, of the Instituto de Astrofisica de Canarias.
The Pan-STARRS1 Surveys (PS1) and the PS1 public science archive have been made possible through contributions by the Institute for Astronomy, the University of Hawaii, the Pan-STARRS Project Office, the Max-Planck Society and its participating institutes, the Max Planck Institute for Astronomy, Heidelberg and the Max Planck Institute for Extraterrestrial Physics, Garching, The Johns Hopkins University, Durham University, the University of Edinburgh, the Queen's University Belfast, the Harvard-Smithsonian Center for Astrophysics, the Las Cumbres Observatory Global Telescope Network Incorporated, the National Central University of Taiwan, the Space Telescope Science Institute, the National Aeronautics and Space Administration under Grant No. NNX08AR22G issued through the Planetary Science Division of the NASA Science Mission Directorate, the National Science Foundation Grant No. AST-1238877, the University of Maryland, Eotvos Lorand University (ELTE), the Los Alamos National Laboratory, and the Gordon and Betty Moore Foundation. 
Funding for the Sloan Digital Sky Survey V has been provided by the Alfred P. Sloan Foundation, the Heising-Simons Foundation, the National Science Foundation, and the Participating Institutions. SDSS acknowledges support and resources from the Center for High-Performance Computing at the University of Utah. SDSS telescopes are located at Apache Point Observatory, funded by the Astrophysical Research Consortium and operated by New Mexico State University, and at Las Campanas Observatory, operated by the Carnegie Institution for Science. The SDSS web site is \url{www.sdss.org}.
SDSS is managed by the Astrophysical Research Consortium for the Participating Institutions of the SDSS Collaboration, including Caltech, The Carnegie Institution for Science, Chilean National Time Allocation Committee (CNTAC) ratified researchers, The Flatiron Institute, the Gotham Participation Group, Harvard University, Heidelberg University, The Johns Hopkins University, L’Ecole polytechnique f\'ed\'erale de Lausanne (EPFL), Leibniz-Institut f\"ur Astrophysik Potsdam (AIP), Max-Planck-Institut f\"ur Astronomie (MPIA Heidelberg), Max-Planck-Institut f\"ur Extraterrestrische Physik (MPE), Nanjing University, National Astronomical Observatories of China (NAOC), New Mexico State University, The Ohio State University, Pennsylvania State University, Smithsonian Astrophysical Observatory, Space Telescope Science Institute (STScI), the Stellar Astrophysics Participation Group, Universidad Nacional Aut\'{o}noma de M\'exico, University of Arizona, University of Colorado Boulder, University of Illinois at Urbana-Champaign, University of Toronto, University of Utah, University of Virginia, Yale University, and Yunnan University.“
\end{acknowledgements}

\bibliography{pubs}

\begin{thebibliography}{109}
\expandafter\ifx\csname natexlab\endcsname\relax\def\natexlab#1{#1}\fi

\bibitem[{{Abdurro'uf} {et~al.}(2022){Abdurro'uf}, {Accetta}, {Aerts}, {Silva
  Aguirre}, {Ahumada}, {Ajgaonkar}, {Filiz Ak}, {Alam}, {Allende Prieto},
  {Almeida}, {Anders}, {Anderson}, {Andrews}, {Anguiano}, {Aquino-Ort{\'\i}z},
  {Arag{\'o}n-Salamanca}, {Argudo-Fern{\'a}ndez}, {Ata}, {Aubert},
  {Avila-Reese}, {Badenes}, {Barb{\'a}}, {Barger}, {Barrera-Ballesteros},
  {Beaton}, {Beers}, {Belfiore}, {Bender}, {Bernardi}, {Bershady}, {Beutler},
  {Bidin}, {Bird}, {Bizyaev}, {Blanc}, {Blanton}, {Boardman}, {Bolton},
  {Boquien}, {Borissova}, {Bovy}, {Brandt}, {Brown}, {Brownstein}, {Brusa},
  {Buchner}, {Bundy}, {Burchett}, {Bureau}, {Burgasser}, {Cabang}, {Campbell},
  {Cappellari}, {Carlberg}, {Wanderley}, {Carrera}, {Cash}, {Chen}, {Chen},
  {Cherinka}, {Chiappini}, {Choi}, {Chojnowski}, {Chung}, {Clerc}, {Cohen},
  {Comerford}, {Comparat}, {da Costa}, {Covey}, {Crane}, {Cruz-Gonzalez},
  {Culhane}, {Cunha}, {Dai}, {Damke}, {Darling}, {Davidson}, {Davies},
  {Dawson}, {De Lee}, {Diamond-Stanic}, {Cano-D{\'\i}az}, {S{\'a}nchez},
  {Donor}, {Duckworth}, {Dwelly}, {Eisenstein}, {Elsworth}, {Emsellem},
  {Eracleous}, {Escoffier}, {Fan}, {Farr}, {Feng}, {Fern{\'a}ndez-Trincado},
  {Feuillet}, {Filipp}, {Fillingham}, {Frinchaboy}, {Fromenteau}, {Galbany},
  {Garc{\'\i}a}, {Garc{\'\i}a-Hern{\'a}ndez}, {Ge}, {Geisler}, {Gelfand},
  {G{\'e}ron}, {Gibson}, {Goddy}, {Godoy-Rivera}, {Grabowski}, {Green},
  {Greener}, {Grier}, {Griffith}, {Guo}, {Guy}, {Hadjara}, {Harding},
  {Hasselquist}, {Hayes}, {Hearty}, {Hern{\'a}ndez}, {Hill}, {Hogg},
  {Holtzman}, {Horta}, {Hsieh}, {Hsu}, {Hsu}, {Huber}, {Huertas-Company},
  {Hutchinson}, {Hwang}, {Ibarra-Medel}, {Chitham}, {Ilha}, {Imig}, {Jaekle},
  {Jayasinghe}, {Ji}, {Johnson}, {Jones}, {J{\"o}nsson}, {Katkov}, {Khalatyan},
  {Kinemuchi}, {Kisku}, {Knapen}, {Kneib}, {Kollmeier}, {Kong}, {Kounkel},
  {Kreckel}, {Krishnarao}, {Lacerna}, {Lane}, {Langgin}, {Lavender}, {Law},
  {Lazarz}, {Leung}, {Leung}, {Lewis}, {Li}, {Li}, {Lian}, {Liang}, {Lin},
  {Lin}, {Lin}, {Lintott}, {Long}, {Longa-Pe{\~n}a}, {L{\'o}pez-Cob{\'a}},
  {Lu}, {Lundgren}, {Luo}, {Mackereth}, {de la Macorra}, {Mahadevan},
  {Majewski}, {Manchado}, {Mandeville}, {Maraston}, {Margalef-Bentabol},
  {Masseron}, {Masters}, {Mathur}, {McDermid}, {Mckay}, {Merloni},
  {Merrifield}, {Meszaros}, {Miglio}, {Di Mille}, {Minniti}, {Minsley},
  {Monachesi}, {Moon}, {Mosser}, {Mulchaey}, {Muna}, {Mu{\~n}oz}, {Myers},
  {Myers}, {Nadathur}, {Nair}, {Nandra}, {Neumann}, {Newman}, {Nidever},
  {Nikakhtar}, {Nitschelm}, {O'Connell}, {Garma-Oehmichen}, {Luan Souza de
  Oliveira}, {Olney}, {Oravetz}, {Ortigoza-Urdaneta}, {Osorio}, {Otter},
  {Pace}, {Padilla}, {Pan}, {Pan}, {Parikh}, {Parker}, {Peirani}, {Pe{\~n}a
  Ram{\'\i}rez}, {Penny}, {Percival}, {Perez-Fournon}, {Pinsonneault},
  {Poidevin}, {Poovelil}, {Price-Whelan}, {B{\'a}rbara de Andrade Queiroz},
  {Raddick}, {Ray}, {Rembold}, {Riddle}, {Riffel}, {Riffel}, {Rix}, {Robin},
  {Rodr{\'\i}guez-Puebla}, {Roman-Lopes}, {Rom{\'a}n-Z{\'u}{\~n}iga}, {Rose},
  {Ross}, {Rossi}, {Rubin}, {Salvato}, {S{\'a}nchez}, {S{\'a}nchez-Gallego},
  {Sanderson}, {Santana Rojas}, {Sarceno}, {Sarmiento}, {Sayres}, {Sazonova},
  {Schaefer}, {Schiavon}, {Schlegel}, {Schneider}, {Schultheis}, {Schwope},
  {Serenelli}, {Serna}, {Shao}, {Shapiro}, {Sharma}, {Shen}, {Shetrone}, {Shu},
  {Simon}, {Skrutskie}, {Smethurst}, {Smith}, {Sobeck}, {Spoo}, {Sprague},
  {Stark}, {Stassun}, {Steinmetz}, {Stello}, {Stone-Martinez},
  {Storchi-Bergmann}, {Stringfellow}, {Stutz}, {Su}, {Taghizadeh-Popp},
  {Talbot}, {Tayar}, {Telles}, {Teske}, {Thakar}, {Theissen}, {Tkachenko},
  {Thomas}, {Tojeiro}, {Hernandez Toledo}, {Troup}, {Trump}, {Trussler},
  {Turner}, {Tuttle}, {Unda-Sanzana}, {V{\'a}zquez-Mata}, {Valentini},
  {Valenzuela}, {Vargas-Gonz{\'a}lez}, {Vargas-Maga{\~n}a}, {Alfaro},
  {Villanova}, {Vincenzo}, {Wake}, {Warfield}, {Washington}, {Weaver},
  {Weijmans}, {Weinberg}, {Weiss}, {Westfall}, {Wild}, {Wilde}, {Wilson},
  {Wilson}, {Wilson}, {Wolf}, {Wood-Vasey}, {Yan}, {Zamora}, {Zasowski},
  {Zhang}, {Zhao}, {Zheng}, {Zheng}, \& {Zhu}}]{2022ApJS..259...35A}
{Abdurro'uf}, {Accetta}, K., {Aerts}, C., {et~al.} 2022, \apjs, 259, 35

\bibitem[{{An} \& {Baan}(2012)}]{2012ApJ...760...77A}
{An}, T. \& {Baan}, W.~A. 2012, \apj, 760, 77

\bibitem[{{An} {et~al.}(2012){An}, {Wu}, {Yang}, {Taylor}, {Hong}, {Baan},
  {Liu}, {Wang}, {Zhang}, {Wang}, {Chen}, {Cui}, {Hao}, \&
  {Zhu}}]{2012ApJS..198....5A}
{An}, T., {Wu}, F., {Yang}, J., {et~al.} 2012, \apjs, 198, 5

\bibitem[{{Axon} {et~al.}(2000){Axon}, {Capetti}, {Fanti}, {Morganti},
  {Robinson}, \& {Spencer}}]{2000AJ....120.2284A}
{Axon}, D.~J., {Capetti}, A., {Fanti}, R., {et~al.} 2000, \aj, 120, 2284

\bibitem[{{Bansal} {et~al.}(2017){Bansal}, {Taylor}, {Peck}, {Zavala}, \&
  {Romani}}]{2017ApJ...843...14B}
{Bansal}, K., {Taylor}, G.~B., {Peck}, A.~B., {Zavala}, R.~T., \& {Romani},
  R.~W. 2017, \apj, 843, 14

\bibitem[{{Baum} {et~al.}(1990){Baum}, {O'Dea}, {Murphy}, \& {de
  Bruyn}}]{1990A&A...232...19B}
{Baum}, S.~A., {O'Dea}, C.~P., {Murphy}, D.~W., \& {de Bruyn}, A.~G. 1990,
  \aap, 232, 19

\bibitem[{{Bicknell} {et~al.}(1997){Bicknell}, {Dopita}, \&
  {O'Dea}}]{1997ApJ...485..112B}
{Bicknell}, G.~V., {Dopita}, M.~A., \& {O'Dea}, C. P.~O. 1997, \apj, 485, 112

\bibitem[{{Bondi} {et~al.}(1998){Bondi}, {Garrett}, \&
  {Gurvits}}]{1998MNRAS.297..559B}
{Bondi}, M., {Garrett}, M.~A., \& {Gurvits}, L.~I. 1998, \mnras, 297, 559

\bibitem[{{Bronzini} {et~al.}(2024){Bronzini}, {Migliori}, {Vignali},
  {Sobolewska}, {Stawarz}, {Siemiginowska}, {Orienti}, {D'Ammando},
  {Giroletti}, {Principe}, \& {Balasubramaniam}}]{2024A&A...684A..65B}
{Bronzini}, E., {Migliori}, G., {Vignali}, C., {et~al.} 2024, \aap, 684, A65

\bibitem[{{Cassaro}(2020)}]{2020IAUS..342..237C}
{Cassaro}, P. 2020, in Perseus in Sicily: From Black Hole to Cluster Outskirts,
  ed. K.~{Asada}, E.~{de Gouveia Dal Pino}, M.~{Giroletti}, H.~{Nagai}, \&
  R.~{Nemmen}, Vol. 342, 237--238

\bibitem[{{Chambers} {et~al.}(2016){Chambers}, {Magnier}, {Metcalfe},
  {Flewelling}, {Huber}, {Waters}, {Denneau}, {Draper}, {Farrow}, {Finkbeiner},
  {Holmberg}, {Koppenhoefer}, {Price}, {Rest}, {Saglia}, {Schlafly}, {Smartt},
  {Sweeney}, {Wainscoat}, {Burgett}, {Chastel}, {Grav}, {Heasley}, {Hodapp},
  {Jedicke}, {Kaiser}, {Kudritzki}, {Luppino}, {Lupton}, {Monet}, {Morgan},
  {Onaka}, {Shiao}, {Stubbs}, {Tonry}, {White}, {Ba{\~n}ados}, {Bell},
  {Bender}, {Bernard}, {Boegner}, {Boffi}, {Botticella}, {Calamida},
  {Casertano}, {Chen}, {Chen}, {Cole}, {Deacon}, {Frenk}, {Fitzsimmons},
  {Gezari}, {Gibbs}, {Goessl}, {Goggia}, {Gourgue}, {Goldman}, {Grant},
  {Grebel}, {Hambly}, {Hasinger}, {Heavens}, {Heckman}, {Henderson}, {Henning},
  {Holman}, {Hopp}, {Ip}, {Isani}, {Jackson}, {Keyes}, {Koekemoer}, {Kotak},
  {Le}, {Liska}, {Long}, {Lucey}, {Liu}, {Martin}, {Masci}, {McLean}, {Mindel},
  {Misra}, {Morganson}, {Murphy}, {Obaika}, {Narayan}, {Nieto-Santisteban},
  {Norberg}, {Peacock}, {Pier}, {Postman}, {Primak}, {Rae}, {Rai}, {Riess},
  {Riffeser}, {Rix}, {R{\"o}ser}, {Russel}, {Rutz}, {Schilbach}, {Schultz},
  {Scolnic}, {Strolger}, {Szalay}, {Seitz}, {Small}, {Smith}, {Soderblom},
  {Taylor}, {Thomson}, {Taylor}, {Thakar}, {Thiel}, {Thilker}, {Unger},
  {Urata}, {Valenti}, {Wagner}, {Walder}, {Walter}, {Watters}, {Werner},
  {Wood-Vasey}, \& {Wyse}}]{2016arXiv161205560C}
{Chambers}, K.~C., {Magnier}, E.~A., {Metcalfe}, N., {et~al.} 2016, arXiv
  e-prints, arXiv:1612.05560

\bibitem[{{Chen} {et~al.}(2020){Chen}, {Yu}, \& {Lu}}]{2020ApJ...897...86C}
{Chen}, Y., {Yu}, Q., \& {Lu}, Y. 2020, \apj, 897, 86

\bibitem[{{Cotton} {et~al.}(2003){Cotton}, {Dallacasa}, {Fanti}, {Fanti},
  {Foley}, {Schilizzi}, {Spencer}, {Saikia}, \&
  {Garrington}}]{2003PASA...20...12C}
{Cotton}, W.~D., {Dallacasa}, D., {Fanti}, C., {et~al.} 2003, \pasa, 20, 12

\bibitem[{{Czerny} {et~al.}(2009){Czerny}, {Siemiginowska}, {Janiuk},
  {Nikiel-Wroczy{\'n}ski}, \& {Stawarz}}]{2009ApJ...698..840C}
{Czerny}, B., {Siemiginowska}, A., {Janiuk}, A., {Nikiel-Wroczy{\'n}ski}, B.,
  \& {Stawarz}, {\L}. 2009, \apj, 698, 840

\bibitem[{{Dallacasa} {et~al.}(1998){Dallacasa}, {Bondi}, {Alef}, \&
  {Mantovani}}]{1998A&AS..129..219D}
{Dallacasa}, D., {Bondi}, M., {Alef}, W., \& {Mantovani}, F. 1998, \aaps, 129,
  219

\bibitem[{{Dallacasa} {et~al.}(1995){Dallacasa}, {Fanti}, {Fanti}, {Schilizzi},
  \& {Spencer}}]{1995A&A...295...27D}
{Dallacasa}, D., {Fanti}, C., {Fanti}, R., {Schilizzi}, R.~T., \& {Spencer},
  R.~E. 1995, \aap, 295, 27

\bibitem[{{Dallacasa} {et~al.}(2021){Dallacasa}, {Orienti}, {Fanti}, \&
  {Fanti}}]{2021MNRAS.504.2312D}
{Dallacasa}, D., {Orienti}, M., {Fanti}, C., \& {Fanti}, R. 2021, \mnras, 504,
  2312

\bibitem[{{Dallacasa} {et~al.}(2013){Dallacasa}, {Orienti}, {Fanti}, {Fanti},
  \& {Stanghellini}}]{2013MNRAS.433..147D}
{Dallacasa}, D., {Orienti}, M., {Fanti}, C., {Fanti}, R., \& {Stanghellini}, C.
  2013, \mnras, 433, 147

\bibitem[{{Dallacasa} {et~al.}(2000){Dallacasa}, {Stanghellini}, {Centonza}, \&
  {Fanti}}]{2000A&A...363..887D}
{Dallacasa}, D., {Stanghellini}, C., {Centonza}, M., \& {Fanti}, R. 2000, \aap,
  363, 887

\bibitem[{{De Rosa} {et~al.}(2019){De Rosa}, {Vignali}, {Bogdanovi{\'c}},
  {Capelo}, {Charisi}, {Dotti}, {Husemann}, {Lusso}, {Mayer}, {Paragi},
  {Runnoe}, {Sesana}, {Steinborn}, {Bianchi}, {Colpi}, {del Valle}, {Frey},
  {Gab{\'a}nyi}, {Giustini}, {Guainazzi}, {Haiman}, {Herrera Ruiz},
  {Herrero-Illana}, {Iwasawa}, {Komossa}, {Lena}, {Loiseau}, {Perez-Torres},
  {Piconcelli}, \& {Volonteri}}]{2019NewAR..8601525D}
{De Rosa}, A., {Vignali}, C., {Bogdanovi{\'c}}, T., {et~al.} 2019, \nar, 86,
  101525

\bibitem[{{Deane} {et~al.}(2014){Deane}, {Paragi}, {Jarvis}, {Coriat},
  {Bernardi}, {Fender}, {Frey}, {Heywood}, {Kl{\"o}ckner}, {Grainge}, \&
  {Rumsey}}]{2014Natur.511...57D}
{Deane}, R.~P., {Paragi}, Z., {Jarvis}, M.~J., {et~al.} 2014, \nat, 511, 57

\bibitem[{{Emonts} {et~al.}(2016){Emonts}, {Morganti}, {Villar-Mart{\'\i}n},
  {Hodgson}, {Brogt}, {Tadhunter}, {Mahony}, \&
  {Oosterloo}}]{2016A&A...596A..19E}
{Emonts}, B.~H.~C., {Morganti}, R., {Villar-Mart{\'\i}n}, M., {et~al.} 2016,
  \aap, 596, A19

\bibitem[{{Fanaroff} \& {Riley}(1974)}]{FR74}
{Fanaroff}, B.~L. \& {Riley}, J.~M. 1974, \mnras, 167, 31P

\bibitem[{{Fanti} {et~al.}(2004){Fanti}, {Branchesi}, {Cotton}, {Dallacasa},
  {Fanti}, {Gregorini}, {Murgia}, {Stanghellini}, \&
  {Vigotti}}]{2004A&A...427..465F}
{Fanti}, C., {Branchesi}, M., {Cotton}, W.~D., {et~al.} 2004, \aap, 427, 465

\bibitem[{{Fanti} {et~al.}(1995){Fanti}, {Fanti}, {Dallacasa}, {Schilizzi},
  {Spencer}, \& {Stanghellini}}]{1995A&A...302..317F}
{Fanti}, C., {Fanti}, R., {Dallacasa}, D., {et~al.} 1995, \aap, 302, 317

\bibitem[{{Fanti} {et~al.}(2000){Fanti}, {Pozzi}, {Fanti}, {Baum}, {O'Dea},
  {Bremer}, {Dallacasa}, {Falcke}, {de Graauw}, {Marecki}, {Miley},
  {Rottgering}, {Schilizzi}, {Snellen}, {Spencer}, \&
  {Stanghellini}}]{2000A&A...358..499F}
{Fanti}, C., {Pozzi}, F., {Fanti}, R., {et~al.} 2000, \aap, 358, 499

\bibitem[{{Fanti} {et~al.}(1990){Fanti}, {Fanti}, {Schilizzi}, {Spencer}, {Nan
  Rendong}, {Parma}, {van Breugel}, \& {Venturi}}]{1990A&A...231..333F}
{Fanti}, R., {Fanti}, C., {Schilizzi}, R.~T., {et~al.} 1990, \aap, 231, 333

\bibitem[{{Fey} {et~al.}(2002){Fey}, {Boboltz}, {Charlot}, {Fomalont}, {Lanyi},
  {Zhang}, \& {K-Q VLBI Survey Collaboration}}]{2002AAS...201.7611F}
{Fey}, A.~L., {Boboltz}, D.~A., {Charlot}, P., {et~al.} 2002, in American
  Astronomical Society Meeting Abstracts, Vol. 201, American Astronomical
  Society Meeting Abstracts, 76.11

\bibitem[{{Fey} {et~al.}(1996){Fey}, {Clegg}, \&
  {Fomalont}}]{1996ApJS..105..299F}
{Fey}, A.~L., {Clegg}, A.~W., \& {Fomalont}, E.~B. 1996, \apjs, 105, 299

\bibitem[{{Frey} \& {Titov}(2021)}]{2021RNAAS...5...60F}
{Frey}, S. \& {Titov}, O. 2021, Research Notes of the American Astronomical
  Society, 5, 60

\bibitem[{{Gopal-Krishna} {et~al.}(1983){Gopal-Krishna}, {Patnaik}, \&
  {Steppe}}]{1983A&A...123..107G}
{Gopal-Krishna}, {Patnaik}, A.~R., \& {Steppe}, H. 1983, \aap, 123, 107

\bibitem[{{Greisen}(1990)}]{1990apaa.conf..125G}
{Greisen}, E.~W. 1990, in Acquisition, Processing and Archiving of Astronomical
  Images, 125--142

\bibitem[{{Gugliucci} {et~al.}(2005){Gugliucci}, {Taylor}, {Peck}, \&
  {Giroletti}}]{2005ApJ...622..136G}
{Gugliucci}, N.~E., {Taylor}, G.~B., {Peck}, A.~B., \& {Giroletti}, M. 2005,
  \apj, 622, 136

\bibitem[{{Hopkins} {et~al.}(2008){Hopkins}, {Hernquist}, {Cox}, {Dutta}, \&
  {Rothberg}}]{Hopkins2008}
{Hopkins}, P.~F., {Hernquist}, L., {Cox}, T.~J., {Dutta}, S.~N., \& {Rothberg},
  B. 2008, \apj, 679, 156

\bibitem[{{Hopkins} {et~al.}(2009){Hopkins}, {Lauer}, {Cox}, {Hernquist}, \&
  {Kormendy}}]{Hopkins2009}
{Hopkins}, P.~F., {Lauer}, T.~R., {Cox}, T.~J., {Hernquist}, L., \& {Kormendy},
  J. 2009, \apjs, 181, 486

\bibitem[{{Horton} {et~al.}(2020){Horton}, {Krause}, \&
  {Hardcastle}}]{2020MNRAS.499.5765H}
{Horton}, M.~A., {Krause}, M. G.~H., \& {Hardcastle}, M.~J. 2020, \mnras, 499,
  5765

\bibitem[{{Kellermann} {et~al.}(2004){Kellermann}, {Lister}, {Homan},
  {Vermeulen}, {Cohen}, {Ros}, {Kadler}, {Zensus}, \&
  {Kovalev}}]{2004ApJ...609..539K}
{Kellermann}, K.~I., {Lister}, M.~L., {Homan}, D.~C., {et~al.} 2004, \apj, 609,
  539

\bibitem[{{Kiehlmann} {et~al.}(2024{\natexlab{a}}){Kiehlmann}, {Lister},
  {Readhead}, {Liodakis}, {O'Neill}, {Pearson}, {Sheldahl}, {Siemiginowska},
  {Tassis}, {Taylor}, \& {Wilkinson}}]{2024ApJ...961..240K}
{Kiehlmann}, S., {Lister}, M.~L., {Readhead}, A.~C.~S., {et~al.}
  2024{\natexlab{a}}, \apj, 961, 240

\bibitem[{{Kiehlmann} {et~al.}(2024{\natexlab{b}}){Kiehlmann}, {Readhead},
  {O'Neill}, {Wilkinson}, {Lister}, {Liodakis}, {Bruzewski}, {Pavlidou},
  {Pearson}, {Sheldahl}, {Siemiginowska}, {Tassis}, \&
  {Taylor}}]{2024ApJ...961..241K}
{Kiehlmann}, S., {Readhead}, A.~C.~S., {O'Neill}, S., {et~al.}
  2024{\natexlab{b}}, \apj, 961, 241

\bibitem[{{Krause} {et~al.}(2019){Krause}, {Shabala}, {Hardcastle}, {Bicknell},
  {B{\"o}hringer}, {Chon}, {Nawaz}, {Sarzi}, \& {Wagner}}]{2019MNRAS.482..240K}
{Krause}, M. G.~H., {Shabala}, S.~S., {Hardcastle}, M.~J., {et~al.} 2019,
  \mnras, 482, 240

\bibitem[{{Ku{\'z}micz} {et~al.}(2017){Ku{\'z}micz}, {Jamrozy},
  {Kozie{\l}-Wierzbowska}, \& {We{\.z}gowiec}}]{2017MNRAS.471.3806K}
{Ku{\'z}micz}, A., {Jamrozy}, M., {Kozie{\l}-Wierzbowska}, D., \&
  {We{\.z}gowiec}, M. 2017, \mnras, 471, 3806

\bibitem[{{Labiano} {et~al.}(2005){Labiano}, {O'Dea}, {Gelderman}, {de Vries},
  {Axon}, {Barthel}, {Baum}, {Capetti}, {Fanti}, {Koekemoer}, {Morganti}, \&
  {Tadhunter}}]{2005A&A...436..493L}
{Labiano}, A., {O'Dea}, C.~P., {Gelderman}, R., {et~al.} 2005, \aap, 436, 493

\bibitem[{{Labiano} {et~al.}(2006){Labiano}, {Vermeulen}, {Barthel}, {O'Dea},
  {Gallimore}, {Baum}, \& {de Vries}}]{2006A&A...447..481L}
{Labiano}, A., {Vermeulen}, R.~C., {Barthel}, P.~D., {et~al.} 2006, \aap, 447,
  481

\bibitem[{{Lister} {et~al.}(2003){Lister}, {Kellermann}, {Vermeulen}, {Cohen},
  {Zensus}, \& {Ros}}]{2003ApJ...584..135L}
{Lister}, M.~L., {Kellermann}, K.~I., {Vermeulen}, R.~C., {et~al.} 2003, \apj,
  584, 135

\bibitem[{{Lobanov} \& {Roland}(2005)}]{2005A&A...431..831L}
{Lobanov}, A.~P. \& {Roland}, J. 2005, \aap, 431, 831

\bibitem[{{Marecki} {et~al.}(2003){Marecki}, {Barthel}, {Polatidis}, \&
  {Owsianik}}]{2003PASA...20...16M}
{Marecki}, A., {Barthel}, P.~D., {Polatidis}, A., \& {Owsianik}, I. 2003,
  \pasa, 20, 16

\bibitem[{{Meusinger} \& {Mhaskey}(2024)}]{2024A&A...682A..18M}
{Meusinger}, H. \& {Mhaskey}, M. 2024, \aap, 682, A18

\bibitem[{{Morganti} {et~al.}(2013){Morganti}, {Fogasy}, {Paragi}, {Oosterloo},
  \& {Orienti}}]{2013Sci...341.1082M}
{Morganti}, R., {Fogasy}, J., {Paragi}, Z., {Oosterloo}, T., \& {Orienti}, M.
  2013, Science, 341, 1082

\bibitem[{{Morganti} \& {Oosterloo}(2018)}]{2018A&ARv..26....4M}
{Morganti}, R. \& {Oosterloo}, T. 2018, \aapr, 26, 4

\bibitem[{{Morganti} {et~al.}(2004){Morganti}, {Oosterloo}, {Tadhunter},
  {Vermeulen}, {Pihlstr{\"o}m}, {van Moorsel}, \&
  {Wills}}]{2004A&A...424..119M}
{Morganti}, R., {Oosterloo}, T.~A., {Tadhunter}, C.~N., {et~al.} 2004, \aap,
  424, 119

\bibitem[{{Murgia}(2003)}]{2003PASA...20...19M}
{Murgia}, M. 2003, \pasa, 20, 19

\bibitem[{{Nolting} {et~al.}(2023){Nolting}, {Ball}, \&
  {Nguyen}}]{2023ApJ...948...25N}
{Nolting}, C., {Ball}, J., \& {Nguyen}, T.~M. 2023, \apj, 948, 25

\bibitem[{{O'Dea}(1998)}]{1998PASP..110..493O}
{O'Dea}, C.~P. 1998, \pasp, 110, 493

\bibitem[{{O'Dea} \& {Baum}(1997)}]{1997AJ....113..148O}
{O'Dea}, C.~P. \& {Baum}, S.~A. 1997, \aj, 113, 148

\bibitem[{{O'Dea} {et~al.}(1991){O'Dea}, {Baum}, \&
  {Stanghellini}}]{1991ApJ...380...66O}
{O'Dea}, C.~P., {Baum}, S.~A., \& {Stanghellini}, C. 1991, \apj, 380, 66

\bibitem[{{O'Dea} \& {Saikia}(2021)}]{2021A&ARv..29....3O}
{O'Dea}, C.~P. \& {Saikia}, D.~J. 2021, A.\&A. Rev., 29, 3

\bibitem[{{O'Dea} {et~al.}(2017){O'Dea}, {Worrall}, {Tremblay}, {Clarke},
  {Rothberg}, {Baum}, {Christiansen}, {Mullarkey}, {Noel-Storr}, \&
  {Mittal}}]{2017ApJ...851...87O}
{O'Dea}, C.~P., {Worrall}, D.~M., {Tremblay}, G.~R., {et~al.} 2017, \apj, 851,
  87

\bibitem[{{Orienti}(2016)}]{2016AN....337....9O}
{Orienti}, M. 2016, Astronomische Nachrichten, 337, 9

\bibitem[{{Orienti} \& {Dallacasa}(2008)}]{2008A&A...479..409O}
{Orienti}, M. \& {Dallacasa}, D. 2008, \aap, 479, 409

\bibitem[{{Orienti} \& {Dallacasa}(2020)}]{2020MNRAS.499.1340O}
{Orienti}, M. \& {Dallacasa}, D. 2020, \mnras, 499, 1340

\bibitem[{{Orienti} {et~al.}(2007){Orienti}, {Dallacasa}, \&
  {Stanghellini}}]{2007A&A...475..813O}
{Orienti}, M., {Dallacasa}, D., \& {Stanghellini}, C. 2007, \aap, 475, 813

\bibitem[{{Orienti} {et~al.}(2010{\natexlab{a}}){Orienti}, {Dallacasa}, \&
  {Stanghellini}}]{2010MNRAS.408.1075O}
{Orienti}, M., {Dallacasa}, D., \& {Stanghellini}, C. 2010{\natexlab{a}},
  \mnras, 408, 1075

\bibitem[{{Orienti} {et~al.}(2010{\natexlab{b}}){Orienti}, {Murgia}, \&
  {Dallacasa}}]{2010MNRAS.402.1892O}
{Orienti}, M., {Murgia}, M., \& {Dallacasa}, D. 2010{\natexlab{b}}, \mnras,
  402, 1892

\bibitem[{{Ostorero} {et~al.}(2010){Ostorero}, {Moderski}, {Stawarz},
  {Diaferio}, {Kowalska}, {Cheung}, {Kataoka}, {Begelman}, \&
  {Wagner}}]{2010ApJ...715.1071O}
{Ostorero}, L., {Moderski}, R., {Stawarz}, {\L}., {et~al.} 2010, \apj, 715,
  1071

\bibitem[{{Pawlik} {et~al.}(2016){Pawlik}, {Wild}, {Walcher}, {Johansson},
  {Villforth}, {Rowlands}, {Mendez-Abreu}, \& {Hewlett}}]{2016MNRAS.456.3032P}
{Pawlik}, M.~M., {Wild}, V., {Walcher}, C.~J., {et~al.} 2016, \mnras, 456, 3032

\bibitem[{{Peng} {et~al.}(2002){Peng}, {Ho}, {Impey}, \& {Rix}}]{Peng2002}
{Peng}, C.~Y., {Ho}, L.~C., {Impey}, C.~D., \& {Rix}, H.-W. 2002, \aj, 124, 266

\bibitem[{{Polatidis} \& {Conway}(2003)}]{2003PASA...20...69P}
{Polatidis}, A.~G. \& {Conway}, J.~E. 2003, \pasa, 20, 69

\bibitem[{{Principe} {et~al.}(2021){Principe}, {Di Venere}, {Orienti},
  {Migliori}, {D'Ammando}, {Mazziotta}, \& {Giroletti}}]{2021MNRAS.507.4564P}
{Principe}, G., {Di Venere}, L., {Orienti}, M., {et~al.} 2021, \mnras, 507,
  4564

\bibitem[{{Pushkarev} \& {Kovalev}(2012)}]{2012A&A...544A..34P}
{Pushkarev}, A.~B. \& {Kovalev}, Y.~Y. 2012, \aap, 544, A34

\bibitem[{{Readhead} {et~al.}(2024){Readhead}, {Ravi}, {Blandford}, {Sullivan},
  {Somalwar}, {Begelman}, {Birkinshaw}, {Liodakis}, {Lister}, {Pearson},
  {Taylor}, {Wilkinson}, {Globus}, {Kiehlmann}, {Lawrence}, {Murphy},
  {O'Neill}, {Pavlidou}, {Sheldahl}, {Siemiginowska}, \&
  {Tassis}}]{2024ApJ...961..242R}
{Readhead}, A.~C.~S., {Ravi}, V., {Blandford}, R.~D., {et~al.} 2024, \apj, 961,
  242

\bibitem[{{Readhead} {et~al.}(1996{\natexlab{a}}){Readhead}, {Taylor},
  {Pearson}, \& {Wilkinson}}]{1996ApJ...460..634R}
{Readhead}, A.~C.~S., {Taylor}, G.~B., {Pearson}, T.~J., \& {Wilkinson}, P.~N.
  1996{\natexlab{a}}, \apj, 460, 634

\bibitem[{{Readhead} {et~al.}(1996{\natexlab{b}}){Readhead}, {Taylor}, {Xu},
  {Pearson}, {Wilkinson}, \& {Polatidis}}]{1996ApJ...460..612R}
{Readhead}, A.~C.~S., {Taylor}, G.~B., {Xu}, W., {et~al.} 1996{\natexlab{b}},
  \apj, 460, 612

\bibitem[{{Readhead} {et~al.}(1994){Readhead}, {Xu}, {Pearson}, {Wilkinson}, \&
  {Polatidis}}]{1994cers.conf...17R}
{Readhead}, A.~C.~S., {Xu}, W., {Pearson}, T.~J., {Wilkinson}, P.~N., \&
  {Polatidis}, A.~G. 1994, in Compact Extragalactic Radio Sources, ed. J.~A.
  {Zensus} \& K.~I. {Kellermann}, 17

\bibitem[{{Reynolds} \& {Begelman}(1997)}]{1997ApJ...487L.135R}
{Reynolds}, C.~S. \& {Begelman}, M.~C. 1997, \apjl, 487, L135

\bibitem[{{Rossetti} {et~al.}(2008){Rossetti}, {Dallacasa}, {Fanti}, {Fanti},
  \& {Mack}}]{2008A&A...487..865R}
{Rossetti}, A., {Dallacasa}, D., {Fanti}, C., {Fanti}, R., \& {Mack}, K.~H.
  2008, \aap, 487, 865

\bibitem[{{Saikia} {et~al.}(2003){Saikia}, {Jeyakumar}, {Mantovani}, {Salter},
  {Spencer}, {Thomasson}, \& {Wiita}}]{2003PASA...20...50S}
{Saikia}, D.~J., {Jeyakumar}, S., {Mantovani}, F., {et~al.} 2003, \pasa, 20, 50

\bibitem[{{Sanghera} {et~al.}(1995){Sanghera}, {Saikia}, {Luedke}, {Spencer},
  {Foulsham}, {Akujor}, \& {Tzioumis}}]{1995A&A...295..629S}
{Sanghera}, H.~S., {Saikia}, D.~J., {Luedke}, E., {et~al.} 1995, \aap, 295, 629

\bibitem[{{Scheuer}(1982)}]{1982IAUS...97..163S}
{Scheuer}, P.~A.~G. 1982, in Extragalactic Radio Sources, ed. D.~S. {Heeschen}
  \& C.~M. {Wade}, Vol.~97, 163--165

\bibitem[{{Schulz} {et~al.}(2021){Schulz}, {Morganti}, {Nyland}, {Paragi},
  {Mahony}, \& {Oosterloo}}]{2021A&A...647A..63S}
{Schulz}, R., {Morganti}, R., {Nyland}, K., {et~al.} 2021, \aap, 647, A63

\bibitem[{{Scoville} {et~al.}(2000){Scoville}, {Evans}, {Thompson}, {Rieke},
  {Hines}, {Low}, {Dinshaw}, {Surace}, \& {Armus}}]{2000AJ....119..991S}
{Scoville}, N.~Z., {Evans}, A.~S., {Thompson}, R., {et~al.} 2000, \aj, 119, 991

\bibitem[{{Shepherd}(1997)}]{1997ASPC..125...77S}
{Shepherd}, M.~C. 1997, in Astronomical Society of the Pacific Conference
  Series, Vol. 125, Astronomical Data Analysis Software and Systems VI, ed.
  G.~{Hunt} \& H.~{Payne}, 77

\bibitem[{{Siemiginowska} {et~al.}(2005){Siemiginowska}, {Cheung}, {LaMassa},
  {Burke}, {Aldcroft}, {Bechtold}, {Elvis}, \& {Worrall}}]{2005ApJ...632..110S}
{Siemiginowska}, A., {Cheung}, C.~C., {LaMassa}, S., {et~al.} 2005, \apj, 632,
  110

\bibitem[{{Snellen} {et~al.}(2000){Snellen}, {Schilizzi}, {Miley}, {de Bruyn},
  {Bremer}, \& {R{\"o}ttgering}}]{2000MNRAS.319..445S}
{Snellen}, I.~A.~G., {Schilizzi}, R.~T., {Miley}, G.~K., {et~al.} 2000, \mnras,
  319, 445

\bibitem[{{Sobolewska} {et~al.}(2019){Sobolewska}, {Siemiginowska},
  {Guainazzi}, {Hardcastle}, {Migliori}, {Ostorero}, \&
  {Stawarz}}]{2019ApJ...884..166S}
{Sobolewska}, M., {Siemiginowska}, A., {Guainazzi}, M., {et~al.} 2019, \apj,
  884, 166

\bibitem[{{Sobolewska} {et~al.}(2023){Sobolewska}, {Siemiginowska}, {Migliori},
  {Ostorero}, {Stawarz}, \& {Guainazzi}}]{2023ApJ...948...81S}
{Sobolewska}, M., {Siemiginowska}, A., {Migliori}, G., {et~al.} 2023, \apj,
  948, 81

\bibitem[{{Sokolovsky} {et~al.}(2011){Sokolovsky}, {Kovalev}, {Pushkarev},
  {Mimica}, \& {Perucho}}]{2011A&A...535A..24S}
{Sokolovsky}, K.~V., {Kovalev}, Y.~Y., {Pushkarev}, A.~B., {Mimica}, P., \&
  {Perucho}, M. 2011, \aap, 535, A24

\bibitem[{{Stanghellini}(2003)}]{2003PASA...20..118S}
{Stanghellini}, C. 2003, \pasa, 20, 118

\bibitem[{{Stanghellini} {et~al.}(2001){Stanghellini}, {Dallacasa}, {O'Dea},
  {Baum}, {Fanti}, \& {Fanti}}]{2001A&A...377..377S}
{Stanghellini}, C., {Dallacasa}, D., {O'Dea}, C.~P., {et~al.} 2001, \aap, 377,
  377

\bibitem[{{Stanghellini} {et~al.}(2009{\natexlab{a}}){Stanghellini},
  {Dallacasa}, \& {Orienti}}]{2009AN....330..223S}
{Stanghellini}, C., {Dallacasa}, D., \& {Orienti}, M. 2009{\natexlab{a}},
  Astronomische Nachrichten, 330, 223

\bibitem[{{Stanghellini} {et~al.}(2009{\natexlab{b}}){Stanghellini},
  {Dallacasa}, {Venturi}, {An}, \& {Hong}}]{2009AN....330..153S}
{Stanghellini}, C., {Dallacasa}, D., {Venturi}, T., {An}, T., \& {Hong}, X.~Y.
  2009{\natexlab{b}}, Astronomische Nachrichten, 330, 153

\bibitem[{{Stanghellini} {et~al.}(1997){Stanghellini}, {O'Dea}, {Baum},
  {Dallacasa}, {Fanti}, \& {Fanti}}]{1997A&A...325..943S}
{Stanghellini}, C., {O'Dea}, C.~P., {Baum}, S.~A., {et~al.} 1997, \aap, 325,
  943

\bibitem[{{Stanghellini} {et~al.}(1993){Stanghellini}, {O'Dea}, {Baum}, \&
  {Laurikainen}}]{1993ApJS...88....1S}
{Stanghellini}, C., {O'Dea}, C.~P., {Baum}, S.~A., \& {Laurikainen}, E. 1993,
  \apjs, 88, 1

\bibitem[{{Stanghellini} {et~al.}(1998){Stanghellini}, {O'Dea}, {Dallacasa},
  {Baum}, {Fanti}, \& {Fanti}}]{1998A&AS..131..303S}
{Stanghellini}, C., {O'Dea}, C.~P., {Dallacasa}, D., {et~al.} 1998, \aaps, 131,
  303

\bibitem[{{Stanghellini} {et~al.}(2005){Stanghellini}, {O'Dea}, {Dallacasa},
  {Cassaro}, {Baum}, {Fanti}, \& {Fanti}}]{2005A&A...443..891S}
{Stanghellini}, C., {O'Dea}, C.~P., {Dallacasa}, D., {et~al.} 2005, \aap, 443,
  891

\bibitem[{{Stanghellini} {et~al.}(1999){Stanghellini}, {O'Dea}, \&
  {Murphy}}]{1999A&AS..134..309S}
{Stanghellini}, C., {O'Dea}, C.~P., \& {Murphy}, D.~W. 1999, \aaps, 134, 309

\bibitem[{{Sullivan} {et~al.}(2024){Sullivan}, {Blandford}, {Begelman},
  {Birkinshaw}, \& {Readhead}}]{2024MNRAS.528.6302S}
{Sullivan}, A.~G., {Blandford}, R.~D., {Begelman}, M.~C., {Birkinshaw}, M., \&
  {Readhead}, A. C.~S. 2024, \mnras, 528, 6302

\bibitem[{{Taylor} {et~al.}(2000){Taylor}, {Marr}, {Pearson}, \&
  {Readhead}}]{2000ApJ...541..112T}
{Taylor}, G.~B., {Marr}, J.~M., {Pearson}, T.~J., \& {Readhead}, A.~C.~S. 2000,
  \apj, 541, 112

\bibitem[{{Taylor} {et~al.}(1996){Taylor}, {Readhead}, \&
  {Pearson}}]{1996ApJ...463...95T}
{Taylor}, G.~B., {Readhead}, A.~C.~S., \& {Pearson}, T.~J. 1996, \apj, 463, 95

\bibitem[{{Tingay} {et~al.}(2015){Tingay}, {Macquart}, {Collier}, {Rees},
  {Callingham}, {Stevens}, {Carretti}, {Wayth}, {Wong}, {Trott}, {McKinley},
  {Bernardi}, {Bowman}, {Briggs}, {Cappallo}, {Corey}, {Deshpande}, {Emrich},
  {Gaensler}, {Goeke}, {Greenhill}, {Hazelton}, {Johnston-Hollitt}, {Kaplan},
  {Kasper}, {Kratzenberg}, {Lonsdale}, {Lynch}, {McWhirter}, {Mitchell},
  {Morales}, {Morgan}, {Oberoi}, {Ord}, {Prabu}, {Rogers}, {Roshi}, {Udaya
  Shankar}, {Srivani}, {Subrahmanyan}, {Waterson}, {Webster}, {Whitney},
  {Williams}, \& {Williams}}]{2015AJ....149...74T}
{Tingay}, S.~J., {Macquart}, J.~P., {Collier}, J.~D., {et~al.} 2015, \aj, 149,
  74

\bibitem[{{Torniainen} {et~al.}(2005){Torniainen}, {Tornikoski},
  {Ter{\"a}sranta}, {Aller}, \& {Aller}}]{2005A&A...435..839T}
{Torniainen}, I., {Tornikoski}, M., {Ter{\"a}sranta}, H., {Aller}, M.~F., \&
  {Aller}, H.~D. 2005, \aap, 435, 839

\bibitem[{{Tremblay} {et~al.}(2008){Tremblay}, {Taylor}, {Helmboldt},
  {Fassnacht}, \& {Pearson}}]{2008ApJ...684..153T}
{Tremblay}, S.~E., {Taylor}, G.~B., {Helmboldt}, J.~F., {Fassnacht}, C.~D., \&
  {Pearson}, T.~J. 2008, \apj, 684, 153

\bibitem[{{Tremblay} {et~al.}(2016){Tremblay}, {Taylor}, {Ortiz}, {Tremblay},
  {Helmboldt}, \& {Romani}}]{2016MNRAS.459..820T}
{Tremblay}, S.~E., {Taylor}, G.~B., {Ortiz}, A.~A., {et~al.} 2016, \mnras, 459,
  820

\bibitem[{{van Breugel} {et~al.}(1984){van Breugel}, {Miley}, \&
  {Heckman}}]{1984AJ.....89....5V}
{van Breugel}, W., {Miley}, G., \& {Heckman}, T. 1984, \aj, 89, 5

\bibitem[{{Vink} {et~al.}(2006){Vink}, {Snellen}, {Mack}, \&
  {Schilizzi}}]{2006MNRAS.367..928V}
{Vink}, J., {Snellen}, I., {Mack}, K.-H., \& {Schilizzi}, R. 2006, \mnras, 367,
  928

\bibitem[{{Wilkinson} {et~al.}(1994){Wilkinson}, {Polatidis}, {Readhead}, {Xu},
  \& {Pearson}}]{1994ApJ...432L..87W}
{Wilkinson}, P.~N., {Polatidis}, A.~G., {Readhead}, A.~C.~S., {Xu}, W., \&
  {Pearson}, T.~J. 1994, \apjl, 432, L87

\bibitem[{{Wu} {et~al.}(2013){Wu}, {An}, {Baan}, {Hong}, {Stanghellini},
  {Frey}, {Xu}, {Liu}, \& {Wang}}]{2013A&A...550A.113W}
{Wu}, F., {An}, T., {Baan}, W.~A., {et~al.} 2013, \aap, 550, A113

\bibitem[{{Xiang} {et~al.}(2002){Xiang}, {Stanghellini}, {Dallacasa}, \&
  {Haiyan}}]{2002A&A...385..768X}
{Xiang}, L., {Stanghellini}, C., {Dallacasa}, D., \& {Haiyan}, Z. 2002, \aap,
  385, 768

\bibitem[{{Yang} {et~al.}(2017){Yang}, {Liu}, {Yang}, {Mi}, {Cui}, {An},
  {Hong}, \& {Ho}}]{2017MNRAS.471.1873Y}
{Yang}, X., {Liu}, X., {Yang}, J., {et~al.} 2017, \mnras, 471, 1873

\bibitem[{{Yu}(2002)}]{2002MNRAS.331..935Y}
{Yu}, Q. 2002, \mnras, 331, 935

\end{thebibliography}

\begin{appendix}
\onecolumn
\section{Images of the radio sources and their optical hosts}

  \begin{figure*}[h!]
   \centering
   \includegraphics[height=10cm]{0019L.eps}
      \includegraphics[height=10cm]{0019C.eps}  
       \includegraphics[height=10.5cm]{0019X.eps}   
        \includegraphics[height=7cm, bb=0 -150 500 400]{0019Sri.eps} 
   \caption{Radio source 0019-000: VLBA images at 1.67, 4.99, and 8.42 GHz. The bottom right panel shows the optical image from NOT, created by stacking the r and i bands. Here, and in the following radio and optical images, the first contour is set at three times the rms noise level, with contour levels progressing geometrically (-3, 3, 6, 12, 24, 48, etc.). Optical units are arbitrary. A cross in the optical image marks the radio position. The NOT observations are described in \cite{1993ApJS...88....1S}.}
              \label{pic0019}
    \end{figure*}

 \begin{figure*}
   \centering
    \includegraphics[height=9cm]{0108C.eps} 
   \includegraphics[height=9cm]{0108X.eps}  
   \includegraphics[height=9cm]{0108U.eps}   
   \includegraphics[height=8cm, bb=50 50 550 600]{0108Piz2.eps}   
   \caption{Radio source 0108+388: VLBA images at 4.89, 8.30, and 15.36 GHz. The bottom right panel shows the optical image from Pan-STARRS, stacking of i and z bands, superimposed on radio emission in grayscale (VLA: project AS637 at 1.36 GHz; grayscale units are in mJy/beam).}
              \label{pic0108}
    \end{figure*}
    
 \begin{figure*}
   \centering
   \includegraphics[height=11cm]{0316P6alt.eps}  
   \includegraphics[height=11cm]{0316S.eps}     
   \includegraphics[height=12cm]{0316C.eps}      
   \includegraphics[height=6cm, bb=0 -200 500 300]{0316+162Piz.eps}   
   \caption{Radio source 0316+162: VLBA images at 0.609, 2.29, and 4.98 GHz. The bottom right panel shows the optical image from Pan-STARSS, stacking of i and z bands.}
              \label{pic0316}
    \end{figure*}  
    
    \begin{figure*}
   \centering
 \includegraphics[height=9.3cm]{0428P.eps} 
 \includegraphics[height=9.3cm]{0428L.eps} 
  \includegraphics[height=9.3cm]{0428S.eps}  
\includegraphics[height=9.3cm]{0428Clar.eps}     
\includegraphics[height=9.3cm]{0428X.eps}  
\includegraphics[height=6cm, bb=30 0 470 500]{0428Sri.eps} 
   \caption{Radio source 0428+205: VLBA images at 0.333, 1.16, 2.29, 4.99, and 8.42 GHz. The bottom right panel shows the optical image from SDSS, stacking of r and i bands.}
              \label{pic0428}
    \end{figure*}
    
    \begin{figure*}
   \centering
  \includegraphics[height=11cm]{0500X.eps}   
  \includegraphics[height=11cm]{0500U.eps} 
   \includegraphics[height=11.5cm]{0500K.eps}  
 \includegraphics[height=6.5cm, bb=0 -150 500 350]{0500Pi.eps}     
   \caption{Radio source 0500+019: VLBA images at 8.42, 15.37, and 22.23 GHz. The bottom right panel shows the optical image from Pan-STARSS, i band. }
              \label{pic0500}
    \end{figure*}

 \begin{figure*}
   \centering
      \includegraphics[height=11cm]{0710bo30L.eps}
   \includegraphics[height=11cm]{0710bo30C.eps}
    \includegraphics[height=11cm]{0710bo30X.eps}  
     \includegraphics[height=7cm, bb=20 0 520 500]{0710Piz.eps}
   \caption{Radio source 0710+439: VLBA images at 1.64, 4.89, and 8.30 GHz. The bottom right panel shows the optical image from Pan-STARSS, stacking of i and z bands.}
              \label{pic0710}
    \end{figure*}  
    
    \begin{figure*}
   \centering
       \includegraphics[height=10cm]{0941L14.eps}
       \includegraphics[height=10cm]{0941S.eps}    
       \includegraphics[height=10cm, bb=180 50 530 670]{0941X.eps}           
       \includegraphics[height=6cm, bb=-40 -100 340 360]{0941Nr.eps}
   \caption{Radio source 0941-080: VLBA images at 1.41, 2.29, and 8.30 GHz. The bottom right panel shows the optical image from NOT, r band.}
              \label{pic0941}
    \end{figure*}

 \begin{figure*}
   \centering
   \includegraphics[height=10.5cm]{1031L14.eps}   
   \includegraphics[height=10.5cm]{1031C.eps}
   \includegraphics[height=9.5cm, bb=20 100 620 640]{1031X.eps}   
   \includegraphics[height=6cm, bb=90 -50 540 400]{1031Nr.eps}   
   \caption{Radio source 1031+567: VLBA images at 1.41, 4.95, and 8.30 GHz. The bottom right panel shows the optical image from NOT, r band.}
              \label{pic1031}
    \end{figure*}    
    
    \begin{figure*}
   \centering
   \includegraphics[height=8.2cm]{1117S.eps}  
   \includegraphics[height=8.2cm]{1117C.eps}    
   \includegraphics[height=8cm]{1117Pri.eps} 
   \caption{Radio source 1117+146: VLBA images at 2.27 and 4.99 GHz. The bottom panel shows the optical image from Pan-STARSS, stacking of r and i bands.}
              \label{pic1117}
    \end{figure*}

    \begin{figure*}
   \centering
   \includegraphics[height=8cm]{1245L.eps}   
    \includegraphics[height=6cm, bb=20 0 500 500]{1245Pri.eps} 
   \includegraphics[height=7.5cm]{1245Ce.eps}   
   \includegraphics[height=7.5cm]{1245Cw.eps}  
      \includegraphics[height=7.5cm]{1245Xe.eps}   
   \includegraphics[height=7.5cm]{1245Xw.eps}  
   \caption{Radio source 1245-196: VLBA images at 1.67, 4.99, and 8.42 GHz. To better highlight the details, the eastern and western sides are shown separately at 4.99 GHz (middle panels) and 8.42 GHz (bottom panels). The top right panel shows the optical image from Pan-STARRS, stacking of r and i bands.}
              \label{pic1245}
    \end{figure*}
    
     \begin{figure*}
   \centering
    \includegraphics[height=8.5cm]{1323L.eps}      
    \includegraphics[height=8.5cm]{1323C.eps}   
    \includegraphics[height=8cm]{1323X.eps}       
   \includegraphics[height=8cm]{1323Udb.eps}  
   \includegraphics[height=7cm]{1323Ni.eps}   
   \caption{Radio source 1323+321: VLBA images at 1.67, 4.54, 8.30, and 15.4 GHz. The bottom panel shows the optical image from NOT, i band.}
              \label{pic1323}
    \end{figure*}   
    
    \begin{figure*}
   \centering
   \includegraphics[height=11cm]{1345L.eps}  
    \includegraphics[height=11cm]{1345X.eps} 
    \includegraphics[height=12cm, bb=100 0 500 670]{1345U.eps} 
    \includegraphics[height=7.5cm, bb=50 -180 550 320]{1345NOTr.eps} 
   \caption{Radio source 1345+125: VLBA images at 1.67, 8.42, and 15.36 GHz. The bottom right panel shows the optical image from NOT, r band.}
              \label{pic1345}
    \end{figure*}

     \begin{figure*}
   \centering
    \includegraphics[height=8.5cm]{1358P6.eps}   
    \includegraphics[height=8.5cm]{1358L14.eps}
    \includegraphics[height=8cm]{1358C.eps} 
    \includegraphics[height=8cm]{1358K.eps}     
   \includegraphics[height=7cm]{1358NOTr.eps}   
   \caption{Radio source 1358+624: VLBA images at 0.611, 1.41, 4.98, and 22.22 GHz. The bottom panel shows the optical image from NOT, r band.}
              \label{pic1358}
    \end{figure*}   

 \begin{figure*}
   \centering
   \includegraphics[height=9cm]{1404L.eps}    
    \includegraphics[height=9cm]{1404X.eps}       
    \includegraphics[height=10cm]{1404U.eps}          
       \includegraphics[height=7.5cm, bb=0 -100 640 540]{1404NOTri.eps}  
   \caption{Radio source 1404+286: VLBA images at 1.67, 8.42, and 15.36 GHz. The bottom right panel shows the optical image from NOT, stacking of r and i bands.}
              \label{pic1404}
    \end{figure*}

     \begin{figure*}
   \centering
   \includegraphics[height=11cm]{1518L.eps}      
   \includegraphics[height=11cm]{1518C.eps}   
   \includegraphics[height=11cm]{1518X.eps}    
   \includegraphics[height=7cm, bb=50 -100 500 450]{1518Pri.eps}   
   \caption{Radio source 1518+047: VLBA images 1.67, 4.99, and 8.42 GHz. The bottom right panel shows the optical image from Pan-STARRS, stacking of r and i bands.}
              \label{pic1518}
    \end{figure*}   
     \begin{figure*}
   \centering
   \includegraphics[height=9.3cm]{1607L.eps} 
    \includegraphics[height=9.3cm]{1607S.eps} 
   \includegraphics[height=9.3cm]{1607C.eps} 
    \includegraphics[height=11cm]{1607X.eps}    
   \includegraphics[height=7cm, bb=0 -100 480 380]{1607Ni.eps}   
   \caption{Radio source 1607+268: VLBA images at 1.63, 2.30, 4.99, and 8.42 GHz. The bottom right panel shows the optical image from NOT, i band.}
              \label{pic1607}
    \end{figure*} 
          
  \begin{figure*}
   \centering
   \includegraphics[height=10cm]{2008C.eps}   
      \includegraphics[height=10cm]{2008X.eps}  
       \includegraphics[height=10.5cm, bb=90 0 570 700]{2008U.eps}   
       \includegraphics[height=6.5cm, bb=0 -100 480 380]{2008Ni.eps}
   \caption{Radio source 2008-068: VLBA images at 4.99, 8.42, and 15.36 GHz. The bottom right panel shows the optical image from NOT, i band.}
              \label{pic2008c}
    \end{figure*}

    \begin{figure*}
   \centering
  \includegraphics[height=11cm]{2128S.eps}     
  \includegraphics[height=11cm]{2128X.eps}   
  \includegraphics[height=12cm, bb=120 0 520 670]{2128U.eps}     
  \includegraphics[height=6cm, bb=40 -300 490 150]{2128Piz.eps}   
   \caption{Radio source 2128+048: VLBA images at 2.30, 8.42, and 15.36 GHz. The bottom right panel shows the optical image from Pan-STARSS, stacking of i and z bands.}
              \label{pic2128}
    \end{figure*} 
    
     \begin{figure*}
   \centering
   \includegraphics[height=8.5cm]{2210L.eps}  
   \includegraphics[height=8.5cm]{2210C.eps}
   \includegraphics[height=8.5cm, bb=80 100 630 550]{2210X.eps}
   \includegraphics[height=6cm, bb=100 -50 550 400]{2210Sri.eps}   
   \caption{Radio source 2210+016: VLBA images at 1.67, 4.99, and 8.42 GHz. The bottom right panel shows the optical image from SDSS, stacking of r and i bands.}
              \label{pic2210}
    \end{figure*}    

     \begin{figure*}
   \centering
      \includegraphics[height=9.5cm]{2342L2.eps}    
       \includegraphics[height=9.5cm]{2342S.eps}     
   \includegraphics[height=6cm, bb=30 0 480 450]{2342Piz.eps}   
   \caption{Radio source 2342+821: VLBA image at 1.29 and 2.28 GHz. The bottom right panel shows the optical image from Pan-STARSS, stacking of i and z bands.}
              \label{pic2342}
    \end{figure*} 

     \begin{figure*}
   \centering
     \includegraphics[height=8.5cm]{2352L14.eps}    
      \includegraphics[height=8.5cm]{2352S.eps}        
     \includegraphics[height=8.5cm]{2352Cbo30.eps}  
     \includegraphics[height=8.5cm]{2352Xbo30.eps}       
   \includegraphics[height=6.5cm]{2352Priz.eps}   
   \caption{Radio source 2352+495: VLBA images at 1.41, 2.29, 4.89, and 8.30 GHz. The bottom panel shows the optical image from Pan-STARSS, stacking of r, i, and z bands.}
              \label{pic2352}
    \end{figure*} 

\FloatBarrier

\section{Flux densities of components}
    
\begin{longtable}{ccccccccccc}
      \caption{Flux densities of components at different radio bands.}   
      \\
\label{comp}
\endfirsthead
\caption{continued.}\\
\hline
\endhead
\noalign{\smallskip\smallskip}
\noalign{\textbf{Notes.} Exact frequency is indicated for each dataset, as it may vary within the same band for different projects. The spectral index is calculated using the two highest available frequencies. The ratio between VLBA and VLA flux densities was determined at the lower listed frequency, interpolating the VLA flux density from \citet{1998A&AS..131..303S} to match the exact VLBA frequency. An (x) indicates that the component is not detected at this frequency, while (") indicates that the flux density of this component is included in the component indicated above. Flux density errors are discussed in Sect. 3.1.}
\endfoot
\hline
\endlastfoot
\hline\hline
Source& Comp & S$_P$ & S$_L$ & S$_S$ &S$_C$ & S$_X$&S$_U$ & S$_K$&$\alpha_{high}$ &S$_{VLBA}$/S$_{VLA}$   \\
     &       & (mJy) & (mJy)& (mJy)&(mJy) &(mJy) &(mJy) & (mJy)&& \\
\hline
    0019--000&$\nu$\tiny{(GHz)}&&$_{1.67}$&&$_{4.99}$&$_{8.42}$&&&&\\
             &N&...&1677&...&634&367&...&...&1.0&\\    
             &S&...&715&...&231&122&...&...&1.2&\\
             &tot&...&2392&...&865&489&...&...&1.1&0.93\\
            \hline    
    0108+388  &$\nu$\tiny{(GHz)}&&&&$_{4.89}$&$_{8.30}$&$_{15.36}$&&&\\
            &E&...&...&...&766&586&271&...&1.3&\\
            &Ce&...&...&...&x&x&15.1&...&...&\\
            &W&...&...&...&496&326&85.3&...&2.2&\\
            &tot&...&...&...&1262&912&371&...&1.5&0.98\\     
            \hline                          
    0316+162&$\nu$\tiny{(GHz)}&$_{0.609}$&&$_{2.29}$&$_{4.98}$&&&&&\\
             &N&7934&...&4232&1995&...&...&...&1.0&\\
             &Ce1&x&...&x&36.3&...&...&...&...&\\
             &Ce2&x&...&x&6.8&...&...&...&...&\\
             &Ce3&130&...&33.9&9.6&...&...&...&1.6&\\
             &S&1475&...&x&x&...&...&...&...&\\
             &tot&9539&...&4266&2048&...&...&...&0.9&1\\  
            \hline
          0428+205&$\nu$\tiny{(GHz)}&$_{0.332}$&$_{1.16}$&$_{2.29}$&$_{4.99}$&$_{8.42}$&&&&\\
              &N&1105&327&118&46.1&x&...&...&...&\\
             &Ce1&x&x&40.4&26.5&16.9&...&...&0.9&\\
             &Ce2&x&x&119&99.4&61.1&...&...&1.2&\\
             &S&1365&2184&2909&1574&907&...&...&1.1&\\  
            &tot&2470&2511&3186&1746&985&...&...&1.1&0.93\\
            \hline                      
    0500+019&$\nu$\tiny{(GHz)}&&&&&$_{8.42}$&$_{15.36}$&$_{22.23}$&&\\
            &N&...&...&...&...&452&244&192&0.7&\\
            &Ce1&...&...&...&...&679&517&565&-0.2&\\   
            &Ce2&...&...&...&...&234&179&184&-0.1&\\
             &S&...&...&...&...&140&50.9&31.5&1.3&\\
             &tot&...&...&...&...&1505&991&973&0.1&1\\  
            \hline                          
    0710+439&$\nu$\tiny{(GHz)}&&$_{1.64}$&&$_{4.89}$&$_{8.30}$&&&&\\
              &N&...&1280&...&770&436&...&...&1.1&\\
              &Ce&...&434&...&579&389&...&...&0.8& \\
             &S&...&225&...&177&82.5&...&...&1.4&\\
             &tot&...&1939&...&1526&908&...&...&1.0&0.99\\ 
            \hline           
 0941--080&$\nu$\tiny{(GHz)}&&$_{1.41}$&$_{2.29}$&&$_{8.30}$&&&&\\
            &NW&...&1269&792&...&156&...&...&1.3&\\
             &SE&...&1149&744&...&165&...&...&1.2&\\
             &tot&...&2418&1536&...&321&...&...&1.2&0.92\\  
            \hline                          
    1031+567&$\nu$\tiny{(GHz)}&&$_{1.41}$&$_{4.95}$&$_{8.30}$&&&&&\\
            &NE&...&1072&472&323&...&...&...&0.7&\\
            &C&...&x&4.6&2.93&...&...&...&0.9&\\
             &SW&...&772&607&401&...&...&...&0.8&\\
             &tot&...&1844&1084&727&...&...&...&0.8&1\\  
            \hline                          
    1117+146&$\nu$\tiny{(GHz)}&&&$_{2.27}$&$_{4.99}$&&&&&\\
             &NW&...&...&653&355&...&...&...&0.8&\\
             &SE&...&...&845&450&...&...&...&0.8&\\
             &tot&...&...&1498&805&...&...&...&0.8&0.89\\  
            \hline                
    1245--197&$\nu$\tiny{(GHz)}&&$_{1.67}$&&$_{4.99}$&$_{8.42}$&&&&\\
             &W&...&578&...&84.2&40.7&...&...&1.4&\\
             &E&...&3284&...&1521&821&...&...&1.2&\\
             &tot&...&3862&...&1605&862&...&...&1.2&0.83\\    
\hline
\newpage
            \hline    
            \hline
Source& Comp & S$_P$ & S$_L$ & S$_S$ &S$_C$ & S$_X$&S$_U$ & S$_K$&$\alpha_{high}$ &S$_{VLBA}$/S$_{VLA}$   \\
     &       & (mJy) & (mJy)& (mJy)&(mJy) &(mJy) &(mJy) & (mJy)&& \\
            \hline        
1323+321&$\nu$\tiny{(GHz)}&&$_{1.67}$&&$_{4.54}$&$_{8.30}$&$_{15.36}$&&&\\
             &NW&...&2152&...&1310&601&435&...&0.5&\\
             &C&...&x&...&16.7&7.3&9.7&...&-0.5&\\
             &SE&...&1625&...&844&310&191&...&0.8&\\
             &tot&...&3777&...&2171&918&636&...&0.6&0.86\\  
            \hline             
1345+125&$\nu$\tiny{(GHz)}&&$_{1.67}$&&&$_{8.42}$&$_{15.36}$&&&\\
             &N&...&95.2&...&...&36.7&24.8&...&0.7&\\
             &C&...&213&...&...&292&220&...&0.5&\\
             &S&...&4264&...&...&734&360&...&1.2&\\
             &tot&...&4572&...&...&1063&605&...&0.9&0.94\\  
            \hline          
1358+624&$\nu$\tiny{(GHz)}&$_{0.611}$&$_{1.41}$&&$_{4.98}$&&&$_{22.22}$&&\\
             &E&3597&2813&...&983&...&...&142&1.3&\\
             &C&x&x&...&20.1&...&...&32.9&-0.3&\\             
             &W&2290&1560&...&346&...&...&43.8&1.4&\\
             &tot&5887&4373&...&1349&...&...&219&1.2&0.99\\  
            \hline          
1404+286&$\nu$\tiny{(GHz)}&&$_{1.46}$&&&$_{8.42}$&$_{15.36}$&&&\\
             &E&...&829&...&&1815&1005&...&1.0&\\
             &W&...&"&...&&146&57.6&...&1.6&\\
             &Ce&...&x&...&&x&3.7&...&&\\             
             &D&...&29.0&...&&x&x&...&&\\             
             &tot&...&858&...&&1961&1066&...&1.0&0.99\\  
            \hline           
1518+047&$\nu$\tiny{(GHz)}&&$_{1.67}$&&$_{4.99}$&$_{8.42}$&&&&\\
             &N&...&1969&...&392&138&...&...&2.0&\\            
             &S&...&1335&...&479&192&...&...&1.8&\\
             &tot&...&3304&...&871&330&...&...&1.9&0.95\\  
            \hline                
   1607+268&$\nu$\tiny{(GHz)}&...&$_{1.63}$&$_{2.30}$&$_{4.99}$&$_{8.42}$&&&&\\
             &N&...&2401&1846&879&478&...&...&1.2&\\
             &S&...&1841&1524&682&327&...&...&1.4&\\
           &tot&...&4242&3370&1561&805&...&...&1.3&0.96\\  
            \hline         
   2008-068&$\nu$\tiny{(GHz)}&&&&$_{4.99}$&$_{8.42}$&$_{15.36}$&&&\\
            &N&...&...&...&1089&687&366&...&1.1&\\
            &Ce&...&...&...&12.4&9.9&6.7&...&0.7&\\            
            &S&...&...&...&132&68.3&26.9&...&1.6&\\
            &tot&...&...&...&1233&765&400&...&1.1&0.94\\  
            \hline              
2128+048&$\nu$\tiny{(GHz)}&&&$_{2.30}$&&$_{8.42}$&$_{15.36}$&&&\\
    &N&...&...&2272&...&1034&431&...&1.5&\\
    &Ce&...&...&15&...&16&7&...&1.4&\\
    &S&...&...&319&...&67&16&...&2.4&\\
    &tot&...&...&2606&...&1117&454&...&1.5&0.83\\  
            \hline                      
    2210+016&$\nu$\tiny{(GHz)}&&$_{1.67}$&&$_{4.99}$&$_{8.42}$&&&&\\
    &W&...&731&...&222&114&...&...&1.3&\\
             &E&...&1268&...&540&251&...&...&1.5&\\
             &tot&...&1999&...&762&365&...&...&1.4&0.81\\  
            \hline               
2342+821  &$\nu$\tiny{(GHz)}&&$_{1.29}$&$_{2.28}$&&&&&&\\
&W&...&3193&1944&...&...&...&...&0.9&\\
              &Ce&...&182&117&...&...&...&...&0.8&\\             
             &E&...&95&53&...&...&...&...&1.0&\\   
             &tot&...&3470&2114&...&...&...&...&0.9&0.87\\                
            \hline             
    2352+495&$\nu$\tiny{(GHz)}&&$_{1.41}$&$_{2.29}$&$_{4.89}$&$_{8.30}$&&&&\\   
            &N&...&600&397&188&115&...&...&0.9&\\
            &S&...&667&451&206&113&...&...&1.1&\\
            &Ce&...&1172&998&874&702&...&...&0.4&\\
            &C&...&"&20&30&32&...&...&-0.1&\\
            &tot&...&2439&1866&1298&962&...&...&0.6&0.95\\
   \end{longtable}
\end{appendix}

\end{document}